\documentclass[a4paper,twocolumn,11pt,accepted=2020-09-07]{quantumarticle}
\pdfoutput=1

\usepackage[nointlimits]{amsmath}
\usepackage{amssymb, cancel}
\usepackage{exscale}
\usepackage{graphicx} 
\usepackage{color}
\usepackage{stackengine}
\usepackage{txfonts}
\usepackage{algorithm}
\usepackage[noend]{algpseudocode}
\usepackage{mathtools}
\usepackage{hyperref}
\hypersetup{colorlinks=true, citecolor=blue, linkcolor=blue}

\DeclareMathOperator{\tr}{tr}
\DeclareMathOperator{\Tr}{tr}

\newcommand{\bra}[1]{\ensuremath{\left\langle{#1}\right\vert}}
\newcommand{\ket}[1]{\ensuremath{\left\vert{#1}\right\rangle}}
\newcommand{\Bra}[1]{\ensuremath{\big({#1}\big\vert}}
\newcommand{\Ket}[1]{\ensuremath{\big\vert{#1}\big)}}

\newcommand{\ketbra}[2]{\left| #1 \right\rangle\!\!\!\,\left\langle #2 \right|}
\newcommand{\proj}[1]{\ketbra{#1}{#1}}

\newcommand{\res}{\mathsf{R}}
\newcommand{\set}{\mathsf{S}}
\newcommand{\mayeq}{\overset{?}{=}}
\newcommand{\sfx}{\mathsf{X}}
\newcommand{\Layer}{\mathsf{L}}

\newtheorem{theorem}{Theorem}

\usepackage{cleveref} % must be loaded after hyperref
\crefformat{equation}{Eq.~(#2#1#3)} % These change 'equation' to Eq., more PRA-style
\crefmultiformat{equation}{Eqs.~(#2#1#3)}{ and~(#2#1#3)}{, (#2#1#3)}{ and~(#2#1#3)}
\Crefformat{equation}{Equation~(#2#1#3)}
\crefformat{section}{Sec.~#2#1#3}
\Crefformat{section}{Section~#2#1#3}
\crefformat{figure}{Fig.~#2#1#3}
\Crefformat{figure}{Figure~#2#1#3}
\crefformat{aside}{Example~#2#1#3}
\Crefformat{aside}{Example~#2#1#3}
\crefmultiformat{aside}{Examples~(#2#1#3)}{ and~(#2#1#3)}{, (#2#1#3)}{ and~(#2#1#3)}
\Crefmultiformat{aside}{Examples~(#2#1#3)}{ and~(#2#1#3)}{, (#2#1#3)}{ and~(#2#1#3)}
\crefformat{Remark}{Rem.~#2#1#3}
\Crefformat{Remark}{Remark~#2#1#3}

\newcommand{\nn}{\nonumber}
\newcommand{\eg}{\emph{e.g.},~}
\newcommand{\ie}{\emph{i.e.},~}

\definecolor{light-gray}{gray}{0.8}
\makeatletter
\renewcommand*{\@fnsymbol}[1]{\ensuremath{\ifcase#1\or *\or \dagger\or \ddagger\or
   \mathsection\or \mathparagraph\or \|\or **\or \dagger\dagger
   \or \ddagger\ddagger \else\@ctrerr\fi}}
\makeatother

\newcommand{\SNLQPL}{Quantum Performance Laboratory,
Sandia National Laboratories, Albuquerque, NM 87185 and Livermore, CA 94550}

\begin{document}

\title{Detecting crosstalk errors in quantum information processors}
\author{Mohan Sarovar}
\email{mnsarov@sandia.gov}
\author{Timothy Proctor}
\author{Kenneth Rudinger}
\author{Kevin Young}
\author{Erik Nielsen}
\author{Robin Blume-Kohout}

\affiliation{\SNLQPL}

\begin{abstract}
Crosstalk occurs in most quantum computing systems with more than one qubit. It can cause a variety of correlated and nonlocal \emph{crosstalk errors} that can be especially harmful to fault-tolerant quantum error correction, which generally relies on errors being local and relatively predictable.  Mitigating crosstalk errors requires understanding, modeling, and detecting them. In this paper, we introduce a comprehensive framework for crosstalk errors and a protocol for detecting and localizing them.  We give a rigorous definition of crosstalk errors that captures a wide range of disparate physical phenomena that have been called ``crosstalk'', and a concrete model for crosstalk-free quantum processors. Errors that violate this model are crosstalk errors.  Next, we give an equivalent but purely operational (model-independent) definition of crosstalk errors.  Using this definition, we construct a protocol for detecting a large class of crosstalk errors in a multi-qubit processor by finding conditional dependencies between observed experimental probabilities. It is highly efficient, in the sense that the number of unique experiments required scales at most cubically, and very often quadratically, with the number of qubits. We demonstrate the protocol using simulations of 2-qubit and 6-qubit processors.
\end{abstract}

\maketitle

\section{Introduction}
Quantum computing has grown from a theoretical concept into a nascent technology.  Cloud-accessible quantum information processors (QIPs) with 20+ qubits exist today, and ones with around $100$ qubits may appear in the next few years \cite{preskill_quantum_2018}. Fundamental operations -- gates, state preparation and measurements (SPAM) -- are approaching the demanding error rates required by the theory of fault-tolerance on a number of physical platforms, including superconducting qubits and trapped ions \cite{Lad.Jel.etal-2010}. However, as experimentalists and engineers have begun to build systems of 10-20 qubits, it is becoming clear that emergent failure modes may be an even bigger problem than errors in elementary operations.  The most obvious failure mode that emerges at scale is \emph{crosstalk}. 

``Crosstalk" describes a wide range of physical phenomena that vary significantly across physical platforms used for quantum computing.  We will focus, instead, on the visible \emph{effects} of crosstalk on the quantum logical behavior of a physical system that is used and treated like a quantum computer.  We refer to these hardware-agnostic effects as \emph{crosstalk errors} -- deviations from the ideal behavior of quantum gates and circuits, which can be formalized and captured in an architecture-independent way.  Crosstalk errors violate either of two key assumptions that go into any well-behaved model of QIP dynamics: spatial locality, and independence of operations. Gates and other operations are supposed to act non-trivially only in a specific ``target'' region of the QIP, and their action on that region is supposed to be independent of the context in which they are applied.  These assumptions enable tractable models for quantum computing, and crosstalk errors violate them.  Here, we give a rigorous definition of crosstalk errors that captures the effects of crosstalk, while avoiding the need to engage deeply with the physical phenomena themselves.

We begin in \cref{sec:crosstalk_errors} and \cref{sec:defn} by defining what it means for a quantum processor to be ``crosstalk-free'' at the quantum logic level.  In \cref{sec:model}, we construct an explicit error model for Markovian crosstalk-free behavior.  Markovian dynamics that are not consistent with that model constitute crosstalk errors.  Then in \cref{sec:diverse} we discuss the difficulty of detecting arbitrary unknown crosstalk errors and define a class of low-weight crosstalk errors that can be efficiently detected. In \cref{sec:protocol} and \cref{sec:sims}, we take an \emph{operational} approach and show how to detect low-weight crosstalk errors using only correlations between experimental variables -- the settings and the outcomes of experiments. The protocol we develop specifies a set of at most $\mathcal{\tilde{O}}(n^3)$, and often $\mathcal{\tilde{O}}(n^2)$, experiments for detecting crosstalk on an $n$-qubit QIP. The analysis of the data from these experiments uses techniques adapted from causal inference on probabilistic graphical models \cite{koller_probabilities_2009,spirtes_introduction_2010}.

Much recent work has been published on detecting, quantifying, and modeling crosstalk and crosstalk errors in quantum computing devices \cite{bialczak_quantum_2010, neeley_generation_2010, Gambetta:2012ke, piltz_trapped-ion-based_2014, Sheldon:2016fh, mckay_three_2017, debnath_demonstration_2016, heinsoo_rapid_2018, reagor_demonstration_2018, rudinger_probing_2019, proctor_direct_2019, erhard_characterizing_2019, ibmq-device-information_2019, gong_genuine_2019, havlicek_supervised_2019,chen_detector_2019, maciejewski_mitigation_2020}. Variants of Ramsey sequences have been used to detect and quantify coherent coupling between qubits \cite{ibmq-device-information_2019}. This technique is very hardware-specific and typically limited to detecting crosstalk in the form of unwanted Hamiltonian couplings of known form. Several groups have also demonstrated mitigation of crosstalk in readout lines by detailed characterization and compensation \cite{heinsoo_rapid_2018, reagor_demonstration_2018, chen_detector_2019, maciejewski_mitigation_2020} (see also Supplementary Information in Refs. \cite{bialczak_quantum_2010, neeley_generation_2010, debnath_demonstration_2016, gong_genuine_2019, havlicek_supervised_2019}).  A very different approach, which is platform-independent and model-free like the work we present here, is the \emph{simultaneous randomized benchmarking} (SRB) technique for detecting and quantifying crosstalk between pairs of qubits \cite{Gambetta:2012ke,Sheldon:2016fh}. The crosstalk detection protocol we present here is similar in motivation to SRB, and is meant to be used as a light-weight diagnostic for the presence of crosstalk. It is specifically designed to be run efficiently on many-qubit QIPs and identify the crosstalk structure (\ie which qubits have crosstalk errors between them), whereas we are not aware of an application of SRB that reveals crosstalk structure in a many-qubit QIP as efficiently. Moreover, our protocol is designed to detect a wide range of crosstalk errors and is more flexible in terms of the experiments that are performed, allowing it to be tailored towards detection of certain types of crosstalk errors. However, SRB has at least one clear advantage over our protocol; it measures the quantitative \emph{rate} of certain crosstalk errors, whereas our protocol is just designed to detect and localize them, and has limited quantitative ability.  Finally, we note that in a previous paper \cite{rudinger_probing_2019} we gave a protocol for detecting context dependence, including crosstalk, that can be seen as a precursor to the protocol given here.

\section{Crosstalk and crosstalk errors}
\label{sec:crosstalk_errors}

Before we embark on defining things precisely, a brief discussion of exactly \emph{what} we are defining is apropos.  In particular, the distinction between ``crosstalk'' and ``crosstalk errors'' needs further explanation.

Crosstalk is an imprecise but widely used term that appears primarily in electrical engineering and communication theory, and generally refers to ``unwanted coupling between signal paths" \cite{mazda_telecommunications_1993}. 
In experimental quantum computing, the word has been adapted to describe a range of physical phenomena in which some subsystem of an experimental device -- a qubit, field, control line, resonator, photodetector, etc. -- unintentionally affects another subsystem.  

A specific quantum computing device will generally display more than one such effect.  For example, a transmon-based quantum processor might experience
\begin{itemize}
\item Residual coherent couplings between transmons that should be uncoupled,
\item Traditional electromagnetic (EM) crosstalk between microwave lines,
\item Stray on-chip EM fields due to imperfect microwave hygiene,
\item Coupling between readout resonators attached to distinct qubits,
\item 60Hz line noise that influences all the qubits. 
\end{itemize}
Any and all of these phenomena could legitimately be termed crosstalk. All of them are architecture-specific; a trapped-ion processor would have its own endemic crosstalk effects, some analogous to these and some not.

Our goal is to understand and address crosstalk in a platform-independent way that facilitates comparisons between quantum processors without reference to the underlying physics.  This is clearly inconsistent with the established use of the term crosstalk to describe specific physics phenomena.  There is no reasonable direct comparison between an unwanted 2-transmon coupling (measured in MHz) and the intensity of a control laser spillover in a trapped-ion setup (measured in W/m$^2$). But we can legitimately compare their \emph{effects} at the quantum logic level of abstraction, where each device is required to behave like a quantum computer, performing quantum logic gates and quantum circuits.  

We introduce the new term ``crosstalk errors'' for this purpose.  It means any observable effect at the quantum logic level (qubits, gates, quantum circuits, and their associated probabilities) that stems uniquely from some form of physical crosstalk. Some forms of physical crosstalk may result in purely local errors -- \eg independent bit flips -- at the quantum logic level; these are not crosstalk errors (despite their source) because they \emph{could} have been produced by local noise.
Similarly, if physical crosstalk exists but has no effect at the quantum logic level (perhaps because of intentional mitigation) then we say that the system is ``crosstalk-free''.

\section{Definition of crosstalk errors}
\label{sec:defn}

Crosstalk errors are undesired dynamics that violate either (or both) of two principles: locality and independence. In an ideal QIP, each qubit is completely isolated from the rest of the universe, and evolves independently of it, \emph{except} when an operation is applied.  Operations, including gates and measurements, couple qubits to other systems, such as external control fields and/or other qubits.  This coupling is supposed to be precise and limited in scope.  

Unfortunately, real QIPs are not ideal. They experience all manner of noise and errors. Of course, not all errors constitute crosstalk errors. Errors can cause deviations from ideal behavior yet still respect locality and independence.  Unwanted dynamics that \emph{do} violate locality or independence constitute crosstalk errors.  We now make this precise by defining locality and independence.

\noindent\fbox{\parbox{\linewidth}{
\noindent\textbf{Locality of operations}:  A QIP has local operations if and only if the physical implementation of any quantum circuit does not create correlation between any qubits, or disjoint subsets of qubits, unless that circuit contains multiqubit operations that intentionally couple them.
}}

If a processor obeys locality, then it makes sense to talk about the action of operations on their target qubits, and we can go further and define independence.  If locality is violated, then operations do not necessarily have well-defined actions on their targets, and independence may not be well-defined.

\noindent\fbox{\parbox{\linewidth}{
\noindent\textbf{Independence of local operations}:  When an operation (gate, measurement, etc) appears in a quantum circuit acting on target qubits $q$ at time $t$, the dynamical evolution of $q$ at time $t$ does not depend on what other operations (acting on disjoint qubits) appear in the circuit at the same time $t$.
}}

\noindent {\bf Defintion 1:} A QIP's behavior is \emph{crosstalk-free} if its behavior, when implementing arbitrary circuits, satisfies locality and independence.

\section{An explicit error model for crosstalk-free processors}
\label{sec:model}

The definitions in the previous section are abstract.  They neither rely upon nor define a concrete model for crosstalk errors or for crosstalk-free processors.  In this section, we specialize to \emph{Markovian} processors and construct an explicit model for crosstalk-free Markovian processors.  By assuming Markovianity we are able to rule out many conceivable failures and define a model in which only finitely many things can go wrong.  By defining crosstalk-free within this framework, we get a division of Markovian errors into crosstalk-free or \emph{local, independent} errors, and everything else (\ie crosstalk errors).

\subsection{Defining \emph{crosstalk-free} for Markovian QIPs}

\begin{figure}[h!]
\includegraphics[width=\columnwidth]{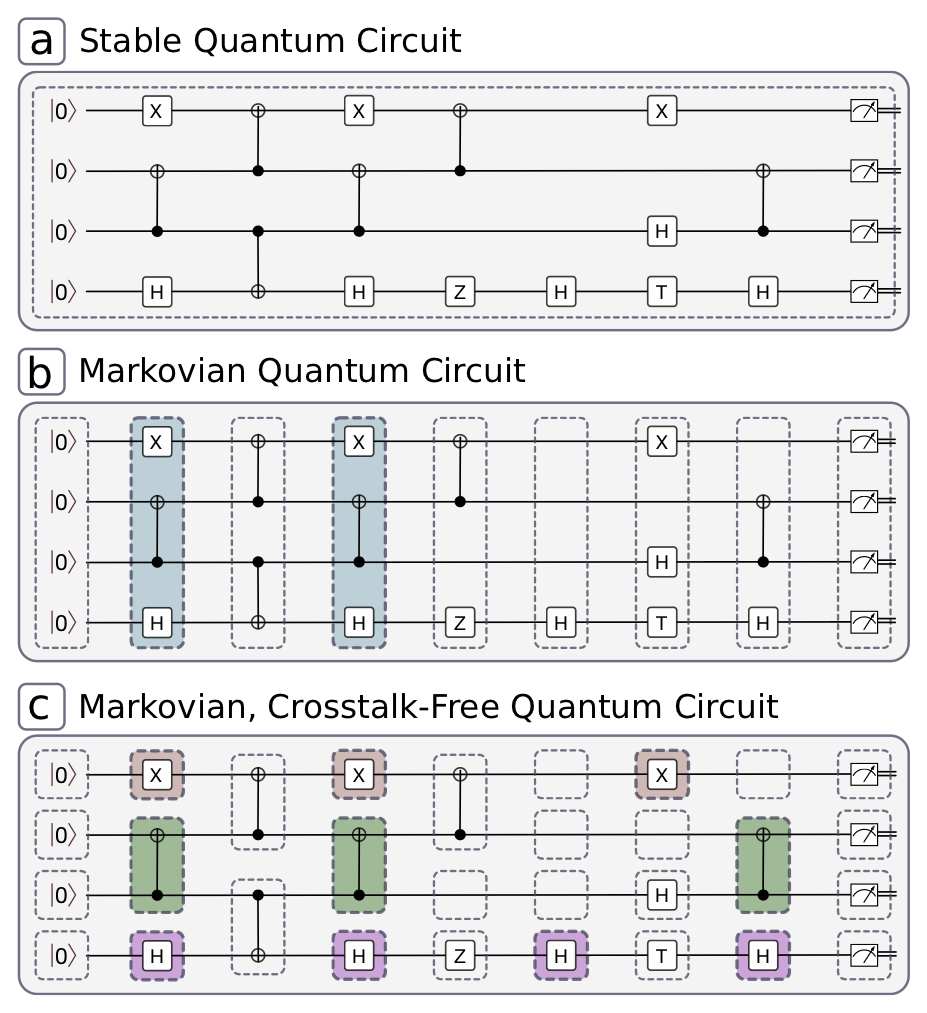}
  \caption{A hierarchy of modularity for QIPs. The dotted lines indicate the modular components in each layer of the hierarchy. (a) A quantum circuit is specified by a schedule of quantum gates on target qubits. It is \emph{stable} if the associated measurement outcome distribution does not depend on any external context. (b) The dynamics of a stable QIP are \emph{Markovian} if each layer in the circuit, including state preparation and measurement, can be represented by a CPTP map that depends only on the operations that comprise that layer, and not on any external context. For example, the two shaded circuit layers are identical and therefore must be represented by the same CPTP map.  (c) The dynamics are Markovian and \emph{crosstalk-free} if the gate operations are modular: the CPTP map describing a given circuit layer can be written as a tensor product of CPTP maps describing each of the component gates (locality), and these component maps do not depend on the other gates in the layer (independence). For example, each appearance of the shaded X, CNOT, or H gates must be represented by the same CPTP maps. \label{fig:markovcrosstalk}}
\end{figure}

We place Markovianity in context within a hierarchy of models for quantum hardware, based on increasing levels of \emph{modularity} (see \cref{fig:markovcrosstalk}): \emph{stable} quantum circuit, \emph{Markovian} quantum circuit, and \emph{Markovian, crosstalk-free} quantum circuit.  
We define each layer in this hierarchy in the following.
%A processor is \emph{stable} if its implementation of any fixed quantum circuit can be described by a fixed distribution over outputs.  It is \emph{Markovian} if each circuit layer is stable and can be described by an $n$-qubit CPTP map.  And a stable, Markovian processor is \emph{crosstalk-free}, if its implementation of quantum operations satisfies locality and independence.

\subsubsection{Stable QIPs}

We call a QIP \emph{stable} if every circuit's outcome probability distribution (over $n$-bit strings) is independent of external contexts \cite{veitia_macroscopic_2017}.  Contexts on which these probabilities might depend include the time at which the circuit is run, the identity of the circuit that was run before it, or even the phase of the moon.  Stability is the weakest notion of modularity: a stable QIP is modular only in the sense that its output distribution is independent of any external contexts, so that each circuit run on the QIP forms a ``module''.  If a QIP is not stable, then modeling or probing its behavior becomes much more difficult. Importantly for this work, protocols for detecting crosstalk will likely be corrupted by this instability, and any results will be unreliable or inconclusive. Fortunately, explicit stability tests for QIPs can often be applied directly to data from other characterization protocols with only minimal modifications to the experiment design. For instance, by repeating a characterization protocol in at least two different contexts, the techniques from Ref.~\cite{rudinger_probing_2019} can verify whether the associated data sets are statistically consistent with one another. To check for the presence of drift,  Ref.~\cite{proctor_detecting_2019} applies Fourier analysis methods to the timestamped output data for each quantum circuit and checks if the resulting spectra are consistent with a stable measurement outcome distribution. Taken together, these approaches can help establish trust that a device is stable, or provide evidence that it is not. 

% The amount of instability can be bounded by using protocols proposed in Refs. \cite{rudinger_probing_2019, proctor_detecting_2019}. 

\subsubsection{Markovian QIPs}

We need a stronger notion of modularity to predict how a QIP will perform on new quantum circuits that have not been run before.  Circuits have a well-defined notion of time, which usually defines a natural division into consecutive \emph{layers} \footnote{The word ``cycle'' (\ie clock cycle) has been used for the same concept. We use ``layer'' to describe a slice of a \emph{circuit}, and reserve ``cycle'' to describe what a specific quantum processor actually \emph{does} in a unit of time.}  of parallel operations (gates, state preparations or measurements).  See \cref{fig:markovcrosstalk} for an example circuit with 9 layers that we notate $\mathsf{L}_0, ..., \mathsf{L}_8$. Operations within a single layer are effectively simultaneous.  A layer is uniquely defined by the list of operations applied to each qubit during that layer, where ``operations'' can include idles, measurements, and initialiation/reset operations as well as elementary gates.  Figure \ref{fig:markovcrosstalk}(b) shows a circuit partitioned into layers. 

We call the QIP \emph{Markovian} if we can describe and model each unique layer by a CPTP map acting on all $n$ qubits in the system.  We use a broad definition of CPTP map here, in which the input and output spaces need not be the same, and can include classical systems.  Typically, an initialization operation is represented by a density matrix, which is a CPTP map from a trivial (1-dimensional) state space to a $d^2$-dimensional vector in quantum state space (Hilbert-Schmidt space); here $d=2^n$.  Elementary gates are represented by ``square'' CPTP maps from a quantum state space to itself.  Terminating measurements are represented by POVMs, which are CPTP maps that map a $d^2$-dimensional quantum state to a $d$-outcome classical distribution.  Intermediate measurements are represented by quantum instruments \cite{davies_operational_1970}, which are CPTP maps that map a $d^2$-dimensional quantum state to a $d^2$-dimensional quantum state \emph{and} a $d$-dimensional classical distribution.  Layers involving multiple kinds of operations are represented by CPTP maps whose input and output spaces correspond to the tensor product of the input/output spaces of all the component operations.

There are many ways to violate the Markovian condition.  For example, a layer might appear multiple times in the circuit, and act differently each time.  But if a QIP is Markovian, then the CPTP map representing each layer depends \emph{only} on the identity of the layer, not on any external context (\eg the time, or which layers occurred previously).  Hereafter, we will only consider Markovian QIPs.  The abstract definitions of locality and independence can presumably be instantiated in a non-Markovian model, but since no general model for non-Markovian processors is known we leave this for future work.

We use $\mathsf{L}$ to denote a CPTP map describing a given circuit layer.  To specify \emph{which} layer $\mathsf{L}$ describes, we will either index its position in the circuit or specify its component operations explicitly. For example, in Fig.~\ref{fig:markovcrosstalk}(b), layers 1 and 3 (highlighted) are identical; each involves an $\rm{X}$ gate on qubit 0, a $\rm{CNOT}$ gate from qubit 2 to qubit 1, and a Hadamard gate on qubit 3.  So we denote the CPTP map for this layer by:
\begin{equation}
	\mathsf{L}_1 = \mathsf{L}_3 = \mathsf{L}(\rm{X}_0,\rm{CNOT_{1\leftarrow 2}},\rm{H}_3).
\end{equation} 

The probabilities of the measurement outcomes for a quantum circuit are determined entirely by the CPTP maps describing the circuit's layers.  For a depth-$N$ circuit that begins with an initialization layer $\rho$, ends with a POVM measurement layer $\{\mathsf{M}_{\mathbf{i}}\}$, and includes $N-2$ gate layers in the middle, the probability of the $\mathbf{i}$th possible result is
\begin{equation}
	\mathsf{Pr}(\mathbf{i}) 
		= \Tr\left[
				\mathsf{M}_{\mathbf{i}}\, 
				\mathsf{L}_{N-2}\circ\cdots\circ\mathsf{L}_2\circ\mathsf{L}_1 (\rho) 
				\right].
\end{equation}
Here $\mathbf{i}$ is an $n$-bit string denoting the measurement result.

Markovianity ensures that the QIP's behavior is modular \emph{in time}. It is the layers that are modular; each layer's effect on the QIP's state must be well-defined, only dependent on identity, and independent of temporal or other contexts. 
This is a powerful assumption.  It makes modeling possible -- we can now predict the results of \emph{new} circuits as long as they are composed of layers that we have characterized already.  

But \emph{efficient} modeling of $n$-qubit circuits requires a stronger modularity condition.  Representing every possible layer by an $n$-qubit CPTP map is neither compact nor tractable.  Exponentially many layers need to be described, and each one requires $\mathcal{O}(16^n)$ real numbers.  Even storing that model is impractical for large $n$, and learning it from data becomes infeasible for as few as three qubits.  Stronger modularity assumptions, like the absence of crosstalk errors, enable efficient models like the one we present below.

Although general $n$-qubit Markovian models are intractable to reconstruct, Markovianity (like stability) can be tested.  Published protocols include those in Refs. \cite{PhysRevB.70.195340,PhysRevLett.103.210401,Wallman_2014, Addis_2015,LI20181,1901.00267}.  Violations of the Markovian model -- generally termed \emph{non-Markovianity} -- may result from a number of underlying causes, including time-dependence, persistent bath memories, or even serial context dependence, where the performance of a layer operation is influenced by the layers that immediately precede (or even follow) it due to the finite bandwidth of control pulses.  

We expect that \emph{all} QIPs are at least a little bit non-Markovian, but we also expect that our Markovian model for crosstalk errors will (like the Markovian CPTP map model itself) fail gracefully, and continue to work well for slightly non-Markovian QIPs.  However, our experience is that crosstalk detection protocols (including the one we develop in the second half of the paper) can confuse violations of Markovianity for crosstalk.  So, in practice, it is important to test for non-Markovian effects before (or simultaneously with) testing for crosstalk.

\subsubsection{Crosstalk-free Markovian QIPs}

Whereas Markovianity allows modularity in time, a processor without crosstalk is modular in \emph{space} -- \ie across qubits and regions.  Layer operations can reliably be composed by combining even smaller operations, that act locally and independently.  Earlier, we said that a processor is crosstalk-free if it obeys locality and independence.  If it is also Markovian, then the conditions for locality and independence can be stated as explicit conditions on the CPTP maps describing circuit layers.

Each \emph{ideal} circuit layer defines a locality (tensor product) structure that partitions the qubits into disjoint and uncoupled target subsets.  A Markovian QIP satisfies locality if and only if the CPTP map describing each layer obeys that structure.  (The proof is trivial: for any bipartite system $AB$ and operation $\mathcal{G}_{AB}$, there exists an initial product state $\rho_A\otimes \rho_B$ such that $\mathcal{G}_{AB}[\rho_A\otimes \rho_B]$ is correlated -- \ie \emph{not} a product state -- if and only if $\mathcal{G}_{AB}$ is not a tensor product of operations).

Therefore, a Markovian model obeys locality if and only if each layer can be represented by a tensor product of local operations.  For such a model, independence is well-defined.  A local, Markovian QIP satisfies independence of operations if and only if each local operation (gate, initialization, measurement) is represented by the \emph{same} local CPTP map in every layer where it appears.  (No proof is needed -- this is just a restatement of the definition of independence above in terms of CPTP maps).

If a Markovian model satisfies both of these conditions, then we say it is crosstalk-free.  Its behavior is consistent with the absence of physical crosstalk, and its dynamics contain no crosstalk errors.  Conversely, any violation of these conditions constitutes a crosstalk error.  

If a QIP satisfies Condition 1 (locality), then each layer's CPTP map is a tensor product over the target subsets implied by that layer. The CPTP map for the layer described above, in the example of Markovianity, would be
\begin{align}
\mathsf{L}(\rm{X}_0,\rm{CNOT_{1\leftarrow 2}},\rm{H}_3) &= \nn \\
		\mathcal{G}(\rm{X}_0) \otimes 
		\mathcal{G}(\rm{CNOT_{1\leftarrow 2}})& \otimes
		\mathcal{G}(\rm{H}_3), 
\end{align}
where $\mathcal{G}$, indexed by the gate operation and qubit target, represents a component CPTP map for that gate.

To satisfy Condition 2 (independence), a gate that appears in multiple layers must act identically in each of them. For example, in Fig.~\ref{fig:markovcrosstalk}, layers 1, 3, 5, and 7 all contain a Hadamard gate acting on the fourth qubit. So the CPTP map describing layer 5 then must take the form
\begin{equation}
\mathsf{L}_5 = \mathsf{L}(\rm{H}_3) = 
		\mathcal{G}(\rm{I}_0) \otimes 
		\mathcal{G}(\rm{I}_1) \otimes
		\mathcal{G}(\rm{I}_2) \otimes
		\mathcal{G}(\rm{H}_3),
\end{equation}
where $\mathcal{G}(\rm{H}_3)$ is the \emph{same} local map that appeared in the other layers (although this map does not have to be the same as the same gate on another qubit, \eg $\mathcal{G}(\rm{H}_4)$).

Initialization and measurement operations must obey the same structure. For the four-qubit QIP shown in Fig.~\ref{fig:markovcrosstalk}, this means that:
\begin{align}
	\rho &= \rho_0 \otimes \rho_1 \otimes \rho_2 \otimes \rho_3 \\
	M_{\mathbf{i}} &= M_{0,i_0} \otimes M_{1,i_1} \otimes M_{2,i_2} \otimes M_{3,i_3},
\end{align}
where $M_{j,i_j}$ is the POVM effect operator for outcome $i_j$ on qubit $j$.  If the initial state is correlated, or the output bit on one qubit depends on another qubit's state, then the QIP is not crosstalk-free.

\subsection{Discussion of the crosstalk-free QIP model}
\noindent 
In classical systems, crosstalk usually refers to a signal in one channel influencing the signal on another channel.  For example, inductive coupling between adjacent copper telephone wires may cause a conversation on one line to be heard on another.  Analogous effects occur in QIPs -- laser beams have finite width and may illuminate neighboring ions, superconducting transmission line resonators may capacitively couple to each other, or qubits themselves may interact directly.  These interactions can be modeled by coupling Hamiltonians.  So it is tempting to say that ``crosstalk'' is nothing more than a coupling Hamiltonian, and the complex abstraction that we have introduced is unnecessary.

But this misses three key points.  First, those Hamiltonians appear in low-level device modeling, and are specific to particular physical implementations.  Second, like all low-level device Hamiltonians, they fluctuate in time and with the state of the environment.  Third, the systems that they couple are often ancillary ones -- control wires, ambient fields, etc -- that would not normally appear in an effective description of the processor and its qubits.  Defining, detecting, and modeling crosstalk at this low level is possible -- and even desirable for device physicists -- but not portable across many devices.

We have presented a high-level, hardware-agnostic \emph{effective} model.  This approach is common.  It is present when qubits are described as 2-dimensional Hilbert spaces, when elementary gates are described by CPTP maps, and when errors are modeled as depolarization or $T_1$ processes.  Our model, like all of those techniques, trades the conceptual simplicity of Hamiltonian dynamics on very large system-specific Hilbert spaces for the practical tractability of an effective model on $n$ qubits. The CPTP map formalism strikes a good balance between rigorous, low-level device models and cross-platform, high-level abstraction -- but as a picture of the underlying physics, it is coarse-grained and can sometimes be counterintuitive.

\begin{figure}[t!]
\centering
  \includegraphics[width=7cm, height=3cm]{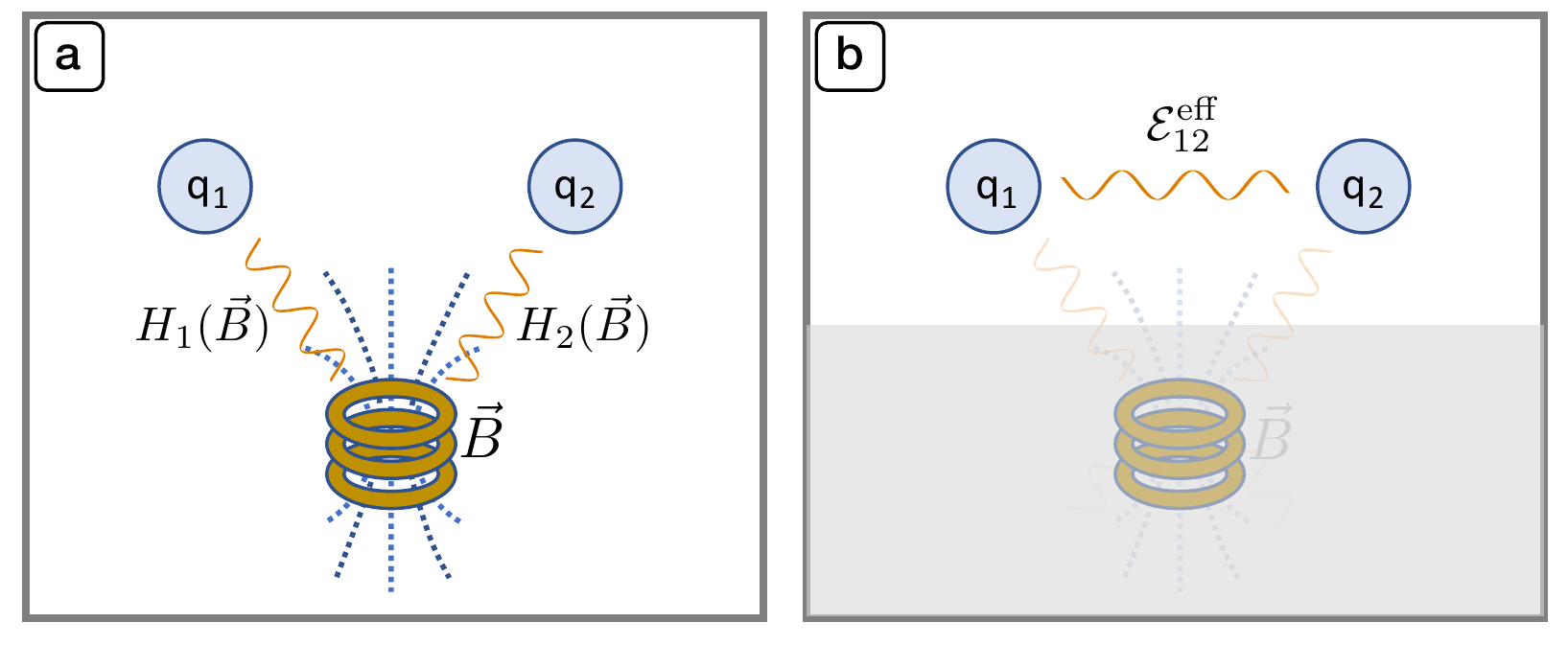}
  \caption{ Two qubits influenced by the same fluctuating magnetic field ($\vec{B}$). {\bf (a)} If the field's state is modeled and tracked, then there is no crosstalk between the two qubits the correlations between their states and errors are fully explained by the field and its coupling to them. {\bf (b)} But if we do not track the field, focusing on the two qubits only then there \emph{is} crosstalk between the qubits, in the form of correlated stochastic errors mediated by the (untracked) magnetic field.   \label{fig:common_B_field}}
\end{figure}

For example, consider two qubits in a magnetic field along the $Z$ axis whose strength varies slowly in time, see \cref{fig:common_B_field}.  The field causes both qubit states to rotate around the $Z$ axis.  Clearly, there is neither coupling nor communication between the qubits. So, if we include the magnetic field in our model, then it seems that there should be no crosstalk between the qubits.  But if we only model the two qubits, and integrate out the field, then the CPTP map describing the \emph{effective} dynamics of the two qubits violates the crosstalk free model -- they experience correlated $Z$ errors, which violate locality.  This may appear counterintuitive, since the qubits are not coupled, and neither has any \emph{causal} effect on the other.  But it reflects the fact that there \emph{is} crosstalk in the system, between each qubit and the magnetic field.  Even when the field is eliminated from the model, it still mediates an effect that creates unexpected correlations between the qubits.  Crosstalk errors can occur at the coarse-grained level even between two qubits that are not directly coupled by the underlying physics.

The stable/Markovian/crosstalk-free hierarchy of models given above is based on strict criteria that, as stated, are either true or false.  One might object that these conditions are practically useless -- no processor is \emph{perfectly} Markovian or crosstalk-free, and could not be proven so even if it were.  While this objection is strictly speaking true, it dismisses the utility of idealized models.  No operation is perfectly unitary, yet unitary dynamics is both well-defined and highly useful \emph{as an ideal}.  In the same way, what matters is not whether a QIP is perfectly crosstalk-free, but how \emph{close} it is to the ideal.  The definitions given above lay the groundwork for metrics that quantify that closeness, and thus for measuring how much crosstalk is present.  

Similarly, perfect Markovianity is not required.  In a real and slightly non-Markovian QIP, we can confidently detect crosstalk as long as the violations of Markovianity (or stability) are small compared to the violations of the crosstalk-free conditions.  An experiment to detect crosstalk has a certain duration and a certain statistical power.  If it detects crosstalk, that conclusion is reliable as long as the QIP's instability and non-Markovianity do not rise above the experiment's level of sensitivity over its duration.

Finally, note that the CPTP maps describing experimental operations are only unique up to a gauge freedom \cite{blume-kohout_demonstration_2017,proctor_what_2017,mavadia_experimental_2018}. In multiqubit QIPs, this gauge freedom is non-local.  Gauge transformations -- which simply change the \emph{description} of the QIP, and have no observable consequences -- can change the tensor product structure of operations, transforming a CPTP map that respects a tensor product structure to one that does not, and vice versa.  This raises the question of whether the ``crosstalk-freeness'' of a model is real and experimentally testable, since it appears to be not gauge-invariant.  

Fortunately, there is a simple resolution:  a stable, Markovian QIP is crosstalk-free if there exists \emph{some} gauge in which Conditions 1-2 hold.  This is directly analogous to the definition of a perfectly error-free gate set.  An ideal target set of operations can be written down in many gauges.  In all but one of them, the CPTP maps \emph{appear} to be different from the original ``ideal'' ones.  But this is the nature of gauge theories.  What matters are the observable probabilities predicted from the theory.  Those are identical in all gauges.  So if there exists \emph{any} gauge in which a gate set coincides with its ideal target, then no experiment (with this gate set) will ever detect any error.  Similarly, if there exists \emph{any} gauge in which a set of $n$-qubit operations is crosstalk-free, then no experiment (with this gate set) will detect evidence of crosstalk. A processor is crosstalk-free if and only if it admits \emph{some} crosstalk-free model.

\subsection{Examples}
\label{sec:egs}

We now consider some examples of crosstalk phenomena, and the crosstalk errors they induce. All the examples in this section involve a QIP with just two qubits, which we label A and B.  The examples can be generalized easily to more qubits.

{\bf \emph{1. Pulse spillover}:} Quantum gates should act only on their target qubits, but control pulses may spill over onto neighboring qubits and affect them.  This is the most widely discussed form of crosstalk, \eg \cite{vandersypen_nmr_2005, piltz_trapped-ion-based_2014, neill_blueprint_2017, buterakos_crosstalk_2018, ma_dissipatively_2019}. For example, consider two qubits that experience no errors when both are idle.  But whenever an $X_{\pi}$ gate is applied to qubit A, the control field spills over onto qubit B and induces a small $X$ rotation.  Each layer still respects the tensor product structure of the two qubits, so locality is not violated.  However, the effect of the idle operation on qubit B depends on whether an idle or an $X_{\pi}$ gate was applied to qubit A at the same time, so this scenario violates independence.

{\bf \emph{2. Always-on Hamiltonian}:} Suppose that when both qubits are idle, they experience an unwanted $XX$ Hamiltonian.  Thus, if A is in the $\ket{+}$ (respectively, $\ket{-}$) state, B undergoes a slow rotation around the $+X$ (respectively, $-X$) axis.  Each qubit is influenced by the state of the other.  
This example violates locality, because the map describing the global idle is an entangling unitary operation, which is not a tensor product of two single-qubit CPTP maps.

{\bf \emph{3. Correlated stochastic errors from common causes}:}  Correlated dynamics caused by a common influence can violate locality. For example (see \cref{fig:common_B_field}), suppose both qubits interact with a common magnetic field along the quantization axis, and that field undergoes white-noise fluctuations.  This produces correlated (weight-2) dephasing or $ZZ$ errors while the qubits are idle.  This is not a tensor product map, and violates locality.  Note that a constant field would only cause local unitary rotations, which respect the tensor product structure and does note result in crosstalk errors.

{\bf \emph{4. Detection crosstalk}:} 
Measurements of a qubit's state may be influenced by the state of neighboring qubits. As an example, consider measuring trapped-ion qubits A and B simultaneously using resonance fluorescence. If light scattered from qubit B spills over onto the detector for qubit A, then the result of measuring qubit A will depend on the state of qubit B. We refer to this type of crosstalk error as detection (or readout) crosstalk, because it specifically affects measurement results. This example violates locality -- the POVM describing the measurement does not respect the QIP's tensor product structure, because the marginal effects corresponding to ``0'' and ``1'' on qubit A act nontrivially on qubit B.

{\bf \emph{5. Correlated state preparation}:} Correlated errors in the controls used to prepare the qubits can create correlated, or even entangled, initial states. This violates locality. For example, consider initializing qubits A and B to the $\ket{0}$ state using a common control field. Occasionally, some noise in the common control field may increase the state preparation error for both qubits. For any single trial, the resulting state would be a product state, but when averaged over many initializations the density matrix describing the initial state can no longer be factorized, so locality is violated.

This list of examples is not exhaustive, but we hope it helps to connect common notions of crosstalk to the conditions that define the crosstalk-free model.

\subsection{Useful terminology for crosstalk errors}

Any violation of the crosstalk-free model results in crosstalk errors, but there are many ways to violate the model.  Some of them are quite distinct from others, both in the physical phenomena that typically produce them, and in their consequences and behavior.  It is useful to identify the most common categories and give them names, if only to facilitate answering the question ``What kind of crosstalk do you see?"  We suggest some useful categories here, based on our experience examining data and modeling noise.  

First, we observe a fundamental difference between errors that violate locality, and those that only violate independence.  Any violation of locality can be traced to at least one specific layer operation that creates unexpected correlations.  We call these crosstalk errors \emph{absolute}.  In contrast, violations of independence \emph{cannot} be isolated to a specific layer operation.  Some local operation just behaves differently in different layers, and no one layer defines the correct behavior of that operation.  We call these crosstalk errors \emph{relative}.

In addition to these terms, which are relatively rigorous, we have found the following less-precise categories to be useful. These categories are not intended to be exhaustive, and may not prove over time to be the most useful classification.  For example, the ``correlated state preparation" example given in the previous section does not fall into any of these categories (it could define another category, but it is not clear that it is sufficiently common or important).  Other violations of the crosstalk-free model can be invented that fall into none of these three categories, or bridge them.  Furthermore, we do not yet have specific protocols for rigorously distinguishing these categories.  Nonetheless, we have found them useful, and so we propose them to the research community.

\textit{Idle crosstalk} is any violation of locality when all qubits are idle.  The unique layer in which no nontrivial operations are performed corresponds to a CPTP map that we call the global idle, and if the global idle is \emph{not} a tensor product of 1-qubit CPTP maps, then we say there is idle crosstalk.  Any error occurring during the global idle that produces correlation between qubits (an error of weight 2 or higher) is an idle crosstalk error.  Examples 2 and 3 in the previous section are examples of idle crosstalk errors.  The same physical phenomena (always-on Hamiltonians, correlated decoherence, \emph{etc.}) can also cause high-weight errors during nontrivial gates, but their effects are usually strongest and easiest to detect during the global idle.

\textit{Operation crosstalk} refers to violations of independence caused by particular elementary operations.  A QIP displays operation crosstalk if the act of performing an operation on qubits in region A changes the dynamics of qubits in a disjoint region B.  It is not always possible to unambiguously ascribe a crosstalk error to an operation (\ie to define operation crosstalk orthogonally to idle crosstalk), but we have found it useful to have terminology for crosstalk errors that change as (non-idle) operations are applied to a QIP. Operation crosstalk is a special case of relative crosstalk, corresponding to cases where the change in region B's dynamics can confidently be blamed on a particular operation.

\textit{Detection crosstalk} refers to violations of locality in the outcomes or results of measurement operations.  If the result of a measurement on one qubit depends on the pre-measurement state of another qubit, that is detection crosstalk.  We avoid the term ``measurement crosstalk" because it is ambiguous; it could also refer to errors on spectator qubits that are caused by measuring a target qubit in the middle of a circuit, which would be an instance of operation crosstalk instead of detection crosstalk.  Example 4 in the previous section is an instance of detection crosstalk.

\section{Crosstalk errors are too diverse to detect without assumptions}
\label{sec:diverse}
Having given a definition of crosstalk errors, we would like to be able to test a QIP to detect their presence, and for further characterization purposes, reveal the structure of crosstalk in the QIP (\ie map out which qubits are most impacted by the crosstalk errors so as to focus the next level of detailed characterization on this subset).
%And if the test reveals crosstalk errors, we would like to learn about their nature -- i.e., to characterize them, both to predict their impact and to help eliminate them.

\subsection{Detecting arbitrary crosstalk errors is hard}

Comprehensive characterization of crosstalk errors is extraordinarily demanding.  Even just detecting any possible crosstalk error is hard (it requires resources that scale super-polynomially with the number of qubits).  Let us demonstrate this.  To begin, we need to define and exclude ``weak'' errors that can be arbitrarily hard to detect.  We say that an experiment $E$ \emph{detects crosstalk} in a stable, Markovian model $M$ if performing $E$ on a QIP described by $M$ has a high probability of producing data that rules out every crosstalk-free model with high confidence.  Crosstalk in a model $M$ is ``strong'' if it can be detected by an experiment using a small number of layers, and ``weak'' if all the experiments that detect it require a large number of layers.  (These concepts are easy to state quantitatively, but it is tedious and not necessary here).

%To formalize this, we define an experiment $E$ as a list of quantum circuits that produce output data $D_E$.  To keep things simple, we list every repeat of every circuit separately -- so if the experiment specifies 10 circuits that are each performed 20 times, then $E$ is a list of 200 circuits.  The \emph{size} of $E$ is the number of layers (including initialization and measurement) required to perform it.  So if each of those 10 circuits involved initialization, 5 layers of gates, and measurement, then the size of the experiment would be $(1+5+1)\times10\times20 = 1400$.  Now, let $M$ be a stable, Markovian model of a QIP that is not crosstalk free, and $E$ be an experiment.  We say that $E$ detects the crosstalk in $M$ at level $\alpha$ if there exists a set of values $\mathcal{D}for the output data $D_E$ such that $Pr( \mathcal{D}|M ) \geq \alpha$, but for every crosstalk-free model $M_0$, $Pr( \mathcal{D}|M_0) \leq 1-\alpha$.  This means that (1) if $E$ is performed and an outcome in $\mathcal{D}$ is observed, we can rule out every crosstalk-free model with high confidence $\alpha$, and (2) if the data are generated by $M$, then an outcome $\mathcal{D}$ will be observed with high probability $\alpha$.  Finally, we can measure the strength of the crosstalk in $M$ by the inverse of the size of the smallest experiment that detects $M$'s crosstalk at reasonably high $\alpha$, e.g. $\alpha = 0.68$ ($\alpha$ can be amplified rapidly toward 1 by just repeating $E$).

Even detecting strong crosstalk errors is hard because crosstalk models $M$ are combinatorially diverse.  Each given one can be detected easily by a tailored experiment, but no small set of experiments can detect them all efficiently.  To illustrate this, we consider two examples, one for relative crosstalk and one for absolute crosstalk.

To see that arbitrary relative crosstalk is hard to detect, consider a QIP that allows any parallel combination of either an $X_{\pi}$ rotation or $I$ (the identity) on each qubit.  Index these possible layers by $n$-bit strings, where ``0'' and ``1'' on the $k$th bit indicate (respectively) that $I$ or $X_{\pi}$ should be performed on the $k$th qubit.  Let $s$ be a randomly selected $n$-bit string, and suppose that every layer \emph{except} the one indexed by $s$ acts perfectly, while applying the one indexed by $s$ depolarizes all qubits.  While each layer respects locality, this model has strong violation of independence because on each qubit there is a gate that causes an error if (and only if) the other $n-1$ qubits are controlled in a particular way.

This crosstalk error is hard to detect because it only occurs if a particular layer (out of exponentially many possible layers) is performed -- but it constitutes strong crosstalk because it is easy to demonstrate by using that layer.

A second example illustrates an analogous problem for absolute crosstalk. Consider the ``idle layer'', where no gates are performed on the qubits. It should act as the $n$-qubit identity channel.  Again, let $s$ be a randomly selected $n$-bit string, and let the idle layer act as the unitary that applies a phase $-1$ to $\ket{s}$ and acts trivially on its complement.  This unitary can easily correlate qubits, so it violates locality.  It is strong, because if $s$ is known, then the correlation can be detected using just a few very short circuits.  But it is also, of course, a Grover oracle for the unknown $s$.  Detecting that it is not the identity is known to be as hard as finding $s$, which requires $O(\sqrt{2^n})$ uses of the layer \cite{beals_quantum_2001}.

This sort of crosstalk is hard to detect because it is very weak on almost all input states.  It only manifests as a significant effect if the input state has high overlap with $\ket{s}$.  So there is a bit of a catch-22:  this crosstalk is strong because it \emph{could} have a dramatic impact on a particular input state, but hard to detect because it has almost no effect on most input states.

Detecting arbitrary strong crosstalk errors is impossible to do efficiently, because it requires testing an exponential number of configurations.  Going further, and characterizing those errors (even partially) is strictly harder.  Designing a protocol to detect crosstalk errors and learn something about them requires specifying something more about the \emph{kind} of errors to be detected, and accepting that other kinds of errors may not be detected.

\subsection{Low-weight crosstalk errors}

We expect characterizing crosstalk in QIPs to require device-specific protocols, informed both by theoretical models of a specific QIP's behavior and the specific tasks or applications that it will run.  But generic protocols are also important.  They provide cross-platform benchmarks, and may detect unexpected errors that tailored protocols miss because of their design.  In the next section, we present a candidate protocol of this type, whose purpose is to (1) detect a significant (but not universal) class of crosstalk errors, and (2) localize those errors, by characterizing which qubits they affect (but not how they act on those qubits).  Since no efficient protocol can be completely generic, some sort of assumptions are necessary to limit the diversity of crosstalk errors.

Our protocol targets \emph{low-weight} crosstalk errors -- ones that result from interactions of just a few subsystems that are supposed to be independent.  In a processor that is crosstalk-free, distinct subsystems \emph{never} interact or develop correlations (note that by ``distinct'' we mean ``not intentionally'' coupled -- two qubits undergoing a 2-qubit gate form a single subsystem for this purpose).  So if the weight of a crosstalk error is (informally) defined as the number of distinct subsystems that it couples together, then all the errors in a crosstalk-free QIP have weight 1.

In contrast, the two examples in the previous subsection illustrated high-weight crosstalk errors.  Each example constructs an input/output function that depends on \emph{all} of its inputs.  In the first example, that function was the map from layer specifications (represented as $n$-bit strings) to CPTP maps.  In the second example, that function was the CPTP map for a single layer, which applied a phase that depended inextricably on every qubit of the input state.  Functions or maps that depend on all their inputs in arbitrary ways are (demonstrably) too diverse, allowing even strong crosstalk to be hidden from efficient detection.

Defining ``weight'' precisely for an absolute crosstalk error, which appears in a specific layer $x$, is straightforward.  That layer's action is represented by a  CPTP map $\Layer(x)$.  Its ideal error-free action can be represented by a CPTP map $\Layer_0(x)$, and so the error in that layer is represented by $\mathcal{E}(x) = \Layer_0(x)^{-1}\Layer(x)$.  The layer's ideal behavior defines a decomposition into distinct subsystems $r_1\otimes r_2 \otimes ...$ that should not interact.  An error map has weight $k$ if it can be written as a convex combination of maps that act trivially (as the identity) on all but $k$ of those subsystems, and cannot be written this way for any smaller $k$.

Typical error maps are not exactly weight-$k$ for any finite $k$ -- e.g., a tensor product of local weak error channels has terms of every weight -- but can be approximated very well by low-weight channels, because the magnitude of the weight-$k$ terms declines exponentially with $k$ above some value.  Henceforth we will take this for granted, and by ``low-weight error map'', we will mean ``error map that can be approximated to high precision by a sum of low-weight terms.''

Quantifying the weight of relative crosstalk errors is slightly more technical.  To do so, we consider a larger state space describing a register of $n$ qubits $\mathcal{Q} = \bigotimes{Q_i}$ \emph{and} a register of $n$ classical digits $\mathcal{C} = \bigotimes{C_i}$. Each $C_i$ specifies what operation is to be performed on the corresponding $Q_i$.  Every possible layer is represented by a distinct state of $\mathcal{C}$, and an entire stable Markovian model can be represented by a \emph{single} operation $\mathcal{M}$ acting on $\mathcal{C}\otimes{Q}$, of the form
\begin{equation}
\mathcal{M} = \sum_{\mathrm{possible ~layer ~specs }~x}{\proj{x}_\mathcal{C}\otimes \Layer(x)_\mathcal{Q}}.
\end{equation}
This is simply a conditional operation, which applies CPTP map $\Layer(x)$ to the qubits, conditional on the classical control register being in state $x$.

Now, as above, we can write $\mathcal{M} = \mathcal{M}_0 \mathcal{E_M}$ so that 
\begin{equation}
\mathcal{E_M} = \sum_{x}{\proj{x}_\mathcal{C}\otimes\mathcal{E}(x)_\mathcal{Q}}
\end{equation}
is the \emph{entire} model's error operation, and perform the same decomposition into weight-$k$ terms.  Now the $i$th subsystem is not just $Q_i$ but $C_i\otimes Q_i$, and a relative crosstalk error that causes $Q_i$ to evolve differently conditional on another qubit's control line $C_j$ is represented by a weight-2 term in $\mathcal{E_M}$
\footnote{We note that it is arguably more elegant to represent error maps including $\mathcal{E_M}$ by their logarithms or generators, and apply the same weight decomposition to them.  The logarithm of a tensor product $\mathcal{E}_1\otimes\mathcal{E}_2$ is a sum of weight-1 terms, $\log(\mathcal{E}_1)\otimes \rm{I} + \rm{I} \otimes \log(\mathcal{E}_2)$. 
So the error generator of a crosstalk-free model is \emph{exactly} weight-1, whereas the error map itself is only approximately weight-1.  However, this representation is less common and requires more machinery that seems unjustified for our purposes here.}. 

Many natural and expected forms of crosstalk have low weight.  Note that low weight does not mean that a single qubit is not perturbed by many other qubits or control lines -- it just means that it is perturbed \emph{independently} by them.  So low-weight crosstalk encompasses many simultaneous few-body interactions.  Moreover, low-weight crosstalk errors are not very diverse.  Simple counting shows that there are only $\mathcal{O}(n)^k$ errors of weight at most $k$ on $n$ qubits, so we can hope to detect any low-weight crosstalk error without devoting exponential resources to the task.  

We do not expect that \emph{all} crosstalk errors will have low weight, but we expect that high-weight errors will stem directly from specific features of the QIP (especially its control architecture, where classical correlations can flourish and induce highly complex dependencies), and are best addressed by tailored, device-specific protocols.  Because low-weight errors are plausible in almost any architecture, a generic protocol to detect and localize them is desirable.

\section{An operational protocol for detecting crosstalk errors}
\label{sec:protocol}

We now return to the abstract definitions of locality and independence presented in \cref{sec:defn} to build a protocol for detecting crosstalk errors, based on the fact that violations of these conditions can be observed directly from operational phenomena.  

%\blue{In theory there are a combinatorial number of ways in which crosstalk errors can enter an $n$-qubit QIP. Therefore, it is an impossible to task to try to detect \emph{any possible} crosstalk error with in an  efficient manner. However, by making reasonable assumptions about the nature of crosstalk error, \eg from locality of the underlying physics generating these errors, we can formulate an efficient detection protocol. }

In \cref{sec:prob_defn} we present the model-free and operational definition of crosstalk-free QIPs that forms the basis of the protocol. In \cref{sec:regions}-\ref{sec:impl} we develop the ingredients of the protocol, including an efficient set of experiments to be performed and tractable data analysis based on statistical tools originally developed for inference on probabilistic graphical models. 
% \cref{sec:expt_design} we discuss the types of experiments that could detect crosstalk in this model-free framework, and design a lightweight protocol that requires only $\mathcal{O}(n)$ circuits for an $n$-qubit QIP. Then in \cref{sec:impl}, we describe how the data from this (and related) protocols can be efficiently analyzed using statistical tools originally developed for inference on probabilistic graphical models. 
In \cref{sec:limitations} we discuss in detail the assumptions behind our protocol and its limitations, especially the crosstalk errors it can and cannot detect. Finally, in \cref{sec:guidance} we present some guidance on how to choose the parameters that define our crosstalk detection protocol based on the physics of the QIP under test.

\subsection{Model-free framework and definitions}
\label{sec:prob_defn}

\begin{figure}[t!]
\centering
  \includegraphics[scale=0.42]{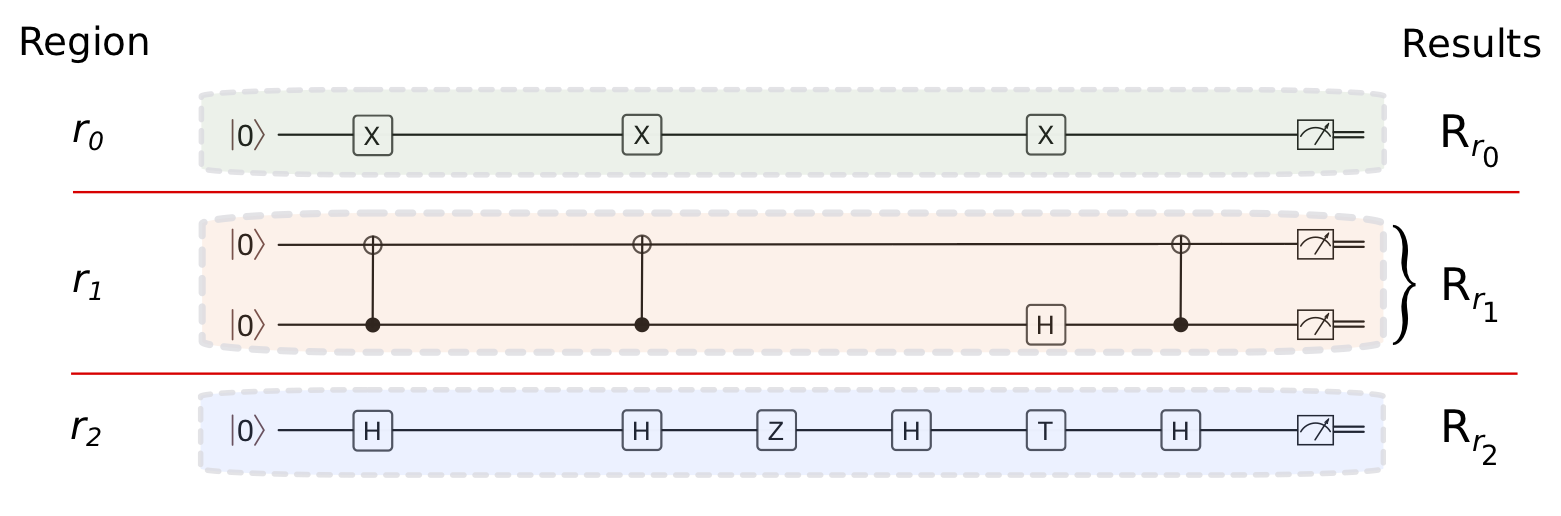}
  \caption{Illustration of the type of circuits used in our protocol. A 4-qubit QIP is partitioned into three regions, labeled $r_0, r_1, r_2$, and the goal is to detect crosstalk errors between these regions. To do so, we perform circuits that \emph{only} apply coupling operations between qubits within a region, never between regions (across the red lines in the figure). The random variable outcome from measuring the qubits in region $r_i$ is denoted $\res_{r_i}$. In this example $\res_{r_0}$ and $\res_{r_2}$ are 1-bit-valued while $\res_{r_1}$ is 2-bit-valued.  \label{fig:regions}}
\end{figure}

Consider a QIP comprising $n$ qubits.  Let $r$ be a partition of the $n$ qubits into $M<n$ disjoint subsets, $r_i \subset \{0, ..., n-1\}$, which we call \emph{regions}, and let $\mathsf{n}(r_i)$ be the number of qubits in region $r_i$.  
We assume no model, only that for each region $r_i$ we (1) apply operations that ideally should only affect qubits in $r_i$ and should not affect qubits in any other region, and (2) make measurements whose results should only depend on the state of qubits in $r_i$.  We will define crosstalk errors in terms of the \emph{settings} that denote the operations applied to a region, and the \emph{results} of measurements on qubits in a region.
An experiment is defined by a tuple $\Omega \equiv (\set_{r_0}, \set_{r_1},...,\set_{r_{M-1}}, \res_{r_0}, \res_{r_1}, ..., \res_{r_{M-1}})$, where $\set_{r_i}$ are the settings assigned to the qubits in region $r_i$ and $\res_{r_i}$ are the measurement results from the qubits in region $r_i$. We treat each member of this tuple as a random variable drawn from some sample space, $\set_{r_i} \in \mathbb{S}_{r_i},\res_{r_i} \in \mathbb{R}_{r_i}$. It is clear that the results are random variables; they are the results of measurements on quantum systems, which are always random variables.  We also treat the settings as random variables, but for a different reason.  In a large QIP, it is not feasible to perform an exhaustive set of experiments that enumerates all the possible experimental settings. So, in practice, observed data constitute a sample over all the possible settings. As we shall see, a random sampling over settings often yields good results.
The random variable $\res_{r_i}$ takes values that are bit strings of length ${\sf n}(r_i)$, obtained by measuring all qubits in region $r_i$ in some basis. More complicated scenarios, \eg involving detection of leakage, are possible but we restrict ourselves to the simplest case here. \cref{fig:regions} illustrates these definitions.

The settings $\set_{r_i}$ are random variables that describe (i) what state is prepared natively on the qubits in $r_i$, (ii) what gates are applied to the qubits in $r_i$, and (iii) what basis the qubits in $r_i$ are measured in. So $\set_{r_i}$ labels a \emph{quantum circuit} for that region (defined here as the state preparation, applied gates and measurement basis choice for a region). 
We note that most quantum computing architectures have only one qubit state that is natively prepared (\eg the ground state) and only one measurement basis (\eg the $Z$ basis). Therefore the only setting that can be varied is the gates applied to the qubits in between state preparation and measurement. Hence in most quantum computing architectures, the settings will be synonymous with ``gates applied to qubits in $r_i$". 

\noindent {\bf Definition 2:} We say that a region $r_i$ is free of crosstalk errors to/from other regions if conditional distributions over the measurement results on this region satisfy:
\begin{align}
	P(\res_{r_i} | \set_{r_i}, \mathsf{T}) &= P(\res_{r_i} | \set_{r_i}), \quad \textrm{with} \nn \\  & \quad \mathsf{T}\subseteq \Omega \setminus \{ \res_{r_i}, \set_{r_i}\} 
	\label{eq:cond_ind}
\end{align}
This means that the distribution of measurement results on region $r_i$ depends \emph{only} on the settings for $r_i$; conditioned on those settings, it is independent of all the other random variables in $\Omega$. Any violation of these conditions is a witness to some kind of crosstalk error.

It is preferable to define the crosstalk-free condition in terms of conditional independence as opposed to \emph{marginal} independence -- \ie $P(\res_{r_i}, \set_{r_j}) = P(\res_{r_i})P( \set_{r_j})$ -- because it is more robust to confounding by hidden (or intentional) correlations in settings, which can become an issue when detecting crosstalk errors in large QIPs. Appendix \ref{sec:condind_marginalind} discusses this further.

This model-free definition of crosstalk errors is equivalent to our model-based definition of crosstalk errors (Definition 1) stated in \cref{sec:defn}; see Appendix \ref{sec:condind_and_physics} for proof. The two definitions capture the same notions of locality and independence of quantum operations -- the model-based definition does so in terms of conditions on models of quantum operations (\ie CPTP maps), while the model-free definition does so in terms of conditions on operational random variables that arise naturally in a QIP. 

\emph{Example.} Here is an example to illustrate the notation introduced above. We wish to detect crosstalk errors induced by single qubit operations on a QIP with 3 qubits, partitioned into two regions $r_0=\{0\}$ and $r_1= \bar{r_0}=\{1,2\}$.  The following elementary single-qubit operations can be performed: initialization in $\ket{0}$; initialization in $\ket{+}$; idle gate (\ie do nothing for one clock cycle); $X_{\pi/2}$ gate; $Z_{\pi/2}$ gate; and measurement in the computational basis. Circuits can be performed that comprise (1) parallel initialization of all 3 qubits, (2) a sequence of $k$ layers built from arbitrary single-qubit gates on each qubit in parallel, and (3) measurement of all qubits in the computational basis. Then the sample space of settings on region $r_0$ -- which includes only qubit 0 -- is
\begin{align}
\mathbb{S}_{r_0} &= \mathbb{S}_{p} \times \mathbb{S}_{g} \nn \\
&= \{\textrm{Prep}_{\ket{0}}, \textrm{Prep}_{\ket{+}}\} \times \{G_I, G_X, G_Z\}^k \nn
\end{align}
where we have distinguished prep settings ($\mathbb{S}_{p}$) and gate settings ($\mathbb{S}_{g}$). 
Only one measurement layer is allowed and the only measurement basis accessible is the computational basis, so there are no measurement settings.  
The space of settings for region $r_1$ is isomorphic to two copies of the settings for $r_0$:  $\mathbb{S}_{r_1} = \mathbb{S}_{r_0} \times \mathbb{S}_{r_0}$. 
The spaces of possible results for each of the two regions are simply $\mathbb{R}_{r_0}=\{0,1\}$ and $\mathbb{R}_{r_1}=\{0,1\}^2$.
In this example, each experiment is labeled by the following tuple of nine random variables,
\begin{align}
	&(\set_{r_0}, \set_{r_1}, \res_{r_0}, \res_{r_1}) \nn \\
	&= ((P_0, G_0), ((P_1, G_1), (P_2, G_2)), R_0, (R_1, R_2)),
\end{align}
where $P_i \in \mathbb{S}_{p}, G_i \in \mathbb{S}_{g}$ and $R_i \in \{0,1\}$ label (respectively) the preparation, sequence of gates, and measurements results for qubit $i$.

The model-free definition given by \cref{eq:cond_ind} leads directly to practical tests for crosstalk, because if we draw a circuit at random from the distribution defined by $P(\set_{r_0}, \set_{r_1},...,\set_{r_{M-1}})$ and perform it on the QIP, the result is a
 sample from the joint probability distribution $P(\Omega) \equiv P(\set_{r_0}, \set_{r_1},...,\set_{r_{M-1}}, \res_{r_0}, \res_{r_1}, ..., \res_{r_{M-1}})$. These samples can be used to statistically test the conditions implied by \cref{eq:cond_ind}. This is, in fact, a general procedure for detecting crosstalk errors. There is always some partitioning of the QIP into regions, some circuit family that can be executed, and some data sample size that will detect \emph{any} crosstalk error using this method. However, as discussed in \cref{sec:diverse}, detecting any possible crosstalk error requires exponential resources, and thus is not a scalable goal. Therefore, our aim is to use this model-independent definition to formulate an efficient protocol that targets low-weight crosstalk errors. 
 
 Developing this protocol requires three ingredients: (i) defining a set of region partitions for a QIP, (ii) defining a set of experiments to perform on the QIP, and (iii) defining an analysis technique on the data produced by these experiments to detect crosstalk using Definition 2. The following subsections tackle each of these ingredients.
  
\subsection{Defining regions}
\label{sec:regions}
Our crosstalk detection protocol looks for correlations between regions of a QIP that should be uncoupled.  This requires partitioning the QIP into disjoint regions.  No single partition into regions will suffice -- for example, we might need to test whether the 2-qubit region $\{1,2\}$ has crosstalk with the 2-qubit region $\{3,4\}$, but also whether $\{2,3\}$ has crosstalk with $\{4,5\}$.  So we need multiple partitions, and for each one, we will define a set of circuits that respect it.

We cannot test every possible partition -- the total number of ways to partition $n$ qubits is $B_n$, the $n$th Bell number, which scales super-exponentially in $n$.  However, testing all possible partitions is unnecessary. Crosstalk errors are associated with individual layers of elementary operations.  In almost every QIP architecture, each elementary operation targets only 1 or 2 qubits.  So, since we focus on low-weight crosstalk errors, it is sufficient to consider partitions into disjoint one- and two-qubit regions.  These allow us to ask (and detect) whether correlations emerge between any two such regions, in circuits that never couple them intentionally.

Let us first set some terminology.
We refer to a region containing exactly $k$ qubits as a \emph{$k$-region}, and a partition of the entire QIP into regions that each contain $k$ or fewer qubits as a \emph{$k$-partition}. We say that a region is \emph{allowed} if it is possible to define circuits that couple all the qubits within that region, without involving any other qubits. So a 2-region is allowed only if the QIP has a 2-qubit gate directly between those qubits.  We say that a region is \emph{in} a given partition if it is one of the regions making up the partition -- \eg region $\{1,2\}$ is in the 6-qubit partition $\{\{1,2\},\{3,4\},\{5,6\}\}$.  Further, a tuple of regions is in a given partition if each region in the tuple is in the partition -- \eg the pair $\{1,2\},\{5,6\}$ is also in the above partition.

There is exactly one unique 1-partition of an $n$-qubit QIP.  So if we wish to detect crosstalk errors associated with single-qubit gates, this is the only partition we need to use.

However, we must also detect crosstalk errors associated with 2-qubit gates, which requires 2-partitions.  The number of possible 2-partitions scales exponentially with $n$; the number of ways to partition $n$ elements into distinct sets of size $k$ (assuming $k$ divides $n$) is $\#(n,k) = \frac{n!}{(k!)^{n/2}(n/k)!}$, and hence $\#(n,2)\approx (\sqrt{n/e})^n$, via the Stirling approximation. This assumes that two-qubit operations are possible between any two qubits in the QIP, however even in the more realistic case of limited connectivity, the number of 2-regions grows exponentially in $n$. So it is impractical to even test all 2-partitions exhaustively.  Fortunately, since we are focused on detecting low-weight crosstalk errors, it suffices to detect pairwise crosstalk between 2-regions (see \cref{sec:limitations} for a discussion of the resulting limitations), and doing this only requires that we guarantee that every pair of 2-regions is in at least one 2-partition of the QIP that is tested.

Since there are at most $n(n-1)/2$ allowed 2-regions, there are $\mathcal{O}(n^4)$ pairs of 2-regions. Therefore, it is easy to define a set of $\mathcal{O}(n^4)$ 2-partitions that contain every such pair (\eg for each of the $\mathcal{O}(n^4)$ possible pairs of 2-regions, define a partition by starting with those two 2-regions and then arbitrarily partitioning the remaining qubits into 2-regions). This requires only poly$(n)$ resources, but is clearly wasteful.  If we are satisfied with a high probability guarantee, a randomized partitioning strategy is more efficient. 

 \begin{theorem}
 	Given an $n$-qubit QIP, let $r_i$, $0<i<R=\frac{n(n-1)}{2}$, be a labeling of all 2-regions in the QIP, and let $\mathcal{P}$ be a set of independent and uniformly sampled random 2-partitions of the QIP. For any  $0<\epsilon<1$, the probability that any pair of distinct 2-regions is in at least one 2-partition is bounded below by $(1-\epsilon)$ if $|\mathcal{P}| \geq n^2(2\log(R)-\log(\epsilon))$. 
 \end{theorem}

\noindent \emph{Proof:} We will follow the logic of the proof of Theorem 3.1 in Ref. \cite{majumdar_why_2018}. Let $p$ be a lower bound on the probability that any 2-partition of the QIP contains a pair of 2-regions (we will see below this bound is the same for any pair). Then, the probability that this pair of 2-regions is not in $|\mathcal{P}|$ random 2-partitions is at most $(1-p)^{|\mathcal{P}|}$. By applying the union bound, we see that the probability that any one of the $R(R-1)/2$ possible pairs of 2-regions in the QIP is not in any partition in $\mathcal{P}$ is at most $\frac{R(R-1)}{2}(1-p)^{|\mathcal{P}|}$. We would like this probability to be at most $\epsilon$, 
\begin{align}
	& \frac{R(R-1)}{2}(1-p)^{|\mathcal{P}|} \leq \epsilon \nn \\
	& \Rightarrow \quad |\mathcal{P}| \geq \frac{\log\left(\frac{\epsilon}{R^2}\right)}{\log(1-p)} \nn \\
	& \Rightarrow \quad |\mathcal{P}| \geq p^{-1}(2\log(R) - \log(\epsilon)), \nn
\end{align}
 where we have used the fact that $-p>\log(1-p)$. It remains to compute the lower bound $p$ in terms of the system parameters. The probability that any pair of 2-regions is in a 2-partition is given by the ratio $\frac{\#(n-4,2)}{\#(n,2)}$ since there are $\#(n-4,2)$ possible partitions of the remaining $n-4$ qubits once the four qubits in the pair of 2-regions have been removed. Computing this ratio, we get $\frac{1}{(n-1)(n-3)}$, and hence a lower bound on this probability is $p>\frac{1}{n^2}$. Substituting this into the above bound on $|\mathcal{P}|$ gives the desired result. $\square$

Either of these approaches -- the brute-force array of $\mathcal{O}(n^4)$ partitions, or the $\mathcal{O}(n^2 \log(n))$ randomized strategy -- defines a list of poly($n$) partitions of the QIP that will detect pairwise crosstalk between any pair of 2-regions with high probability.

We note that in QIPs with local connectivity restrictions, \eg a planar array of qubits where intentional coupling operations are only possible between nearest neighbors qubits, $p^{-1} = \mathcal{O}(n)$, and therefore the scaling of the randomized strategy is improved to $\mathcal{O}(n\log(n))$. Similarly, the scaling of the brute-force partitioning improves to $\mathcal{O}(n^2)$ under local connectivity.
   
\subsection{Lightweight experiment design}
\label{sec:expt_design}
Given a partition of a QIP into regions, we must define a set of circuits to run on the QIP that constitute the crosstalk detection experiment.
We only consider circuits that do not (intentionally) couple regions, which means that for each region there is a well-defined \emph{subcircuit} comprising all operations applied to it.  We also assume, for the sake of simplicity, that the QIP has unique initialization and measurement operations (in $\ket{0}^{\otimes n}$ and the $\{\bra{0},\bra{1}\}^{\otimes n}$ basis).  Thus, the settings for a region correspond precisely to the gates in the subcircuit on that region.

%\red{Our goal here is to detect any kind of crosstalk, between any regions.}
Each possible circuit on the QIP is composed of the parallel application of multiple subcircuits, one on each region.
The simplest approach is to choose a collection of $N_{\rm circ}$ subcircuits for each region, and then perform \emph{all} combinations of those subcircuits. We refer to this collection of subcircuits as the ``bag'' of circuits applied to a region.  (We postpone the question of what subcircuits to place in the bag to the end of this subsection). We call this the  \emph{exhaustive experiment}, in which each subcircuit on region $r_i$ gets performed in an exhaustive variety of different \emph{contexts} -- \ie in parallel with all $N_{\rm circ}$ circuits on other regions -- and so violations of independence are easy to detect in the data.  Unfortunately, this experiment defines a hypercube containing $N_{\rm circ}^M$ distinct circuits, which grows too rapidly with $M$ (the number of regions) to be feasible.

However, we observe that in the exhaustive experiment, each subcircuit on every region $r_i$ is performed in exponentially many distinct contexts (defined by the settings on the other regions $r_j \neq r_i$).  
%This is arduous but powerful -- it can detect contrived and unlikely forms of crosstalk, like a situation where the $M$th region experiences extra errors if and only if a specific subcircuit is performed on \emph{all} of the other regions (\eg if and only if an $X$ gate is performed on $M-1$ qubits, the $M$th qubit experiences additional depolarization).  It is easy to see that detecting arbitrary conditional dynamics of this type demands performing every possible combination of settings. 
%But it is also overkill; real-world crosstalk usually has pairwise manifestations, meaning that if the dynamics of region $r_i$ are correlated with the settings on several regions, they are \emph{also} correlated with each of those regions' settings individually.
This is arduous and overkill; since crosstalk errors are not likely to only be present in one or few of this exponential number of contexts (this is discussed further in \cref{sec:limitations}), we can subsample from this exhaustive experiment.
So we will choose a sparse subset of the experiments in the hypercube defining the exhaustive experiment, with the goal of defining a small set of experiments that allow low-weight crosstalk errors to be detected.  

\subsubsection{An explicit construction}

The sparse sampling of the hypercube should maintain two important properties of the exhaustive experiment. First, each subcircuit in the bag for each region $r_i$ must appear in multiple contexts (but not \emph{exponentially} many).  Second, that set of contexts in which each subcircuit gets performed must vary on \emph{each} of the other regions.  These properties ensure that -- whatever subcircuits we select for each region's bag -- the data will reveal whether the local results of those subcircuits are significantly influenced by the settings (choice of subcircuit) on any other region.

\begin{figure}[t!]
\centering
  \includegraphics[scale=1]{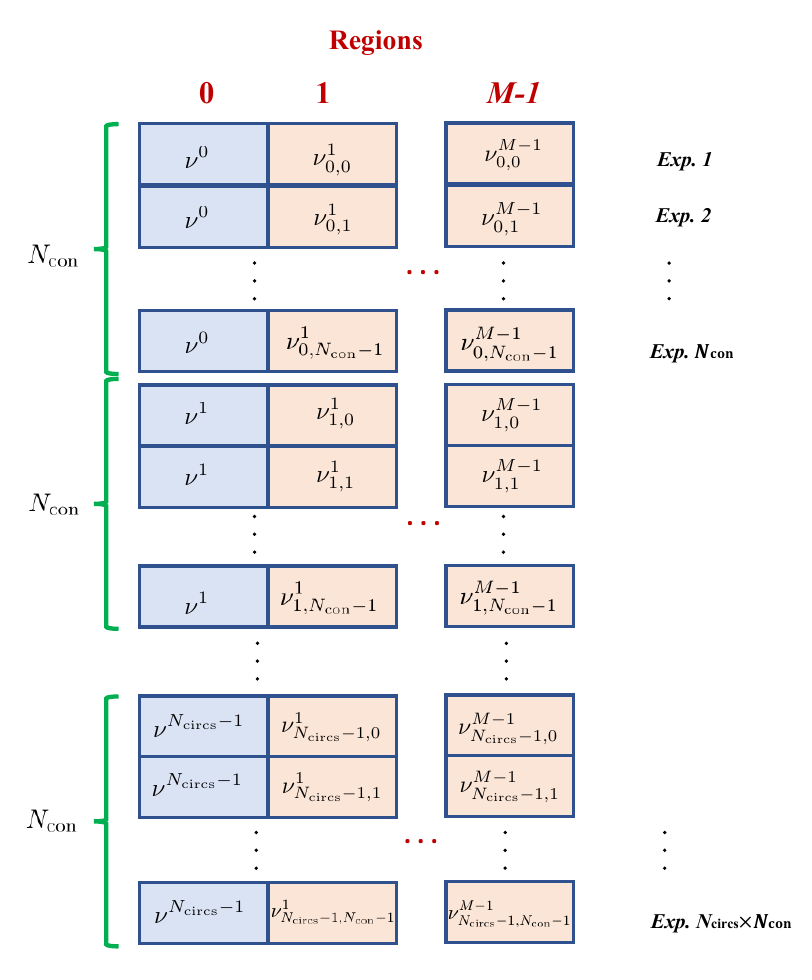}
  \caption{Illustration of the circuits performed in the lightweight crosstalk error detection experiment design. The QIP is partitioned into $M$ non-overlapping regions, and subcircuits of fixed length $L$ are applied to each region. This diagram represents one ``epoch'', during which the $N_{\rm circs}$ subcircuits in the bag for region 0 (denoted $\nu^i$) are iterated over. Each $\nu^i$ is repeated $N_{\rm con}$ times on region 0, and in each instance (each line in the diagram) the subcircuits on the other $M-1$ regions are randomly sampled (with replacement) from the subcircuits in the bag for that region.  In the diagram $\nu^m_{i,j}$ denotes the subcircuit applied to region $m$ in the $j$th context when $\nu^i$ is applied to region 0. The experiment design prescribes $M$ such epochs; in  epoch $m$ the $N_{\rm circs}$ subcircuits in the bag for region $m$ are iterated over while the subcircuits for all the other regions are randomly sampled. It is assumed that all qubits are initialized in their ground states, and the measurements after each of the prescribed circuits are performed simultaneously on all qubits in the computational basis. Finally, each of the experiments is repeated $N_{\rm rep}$ times in order to collect statistics. \label{fig:expts}}
\end{figure}

The construction we outline now ensures that these properties are preserved, even with much fewer experiments. It is defined by three adjustable integer parameters:
\begin{itemize}
\item $L$ is the length or depth of all the subcircuits.  Subcircuits on different regions are applied in parallel, so they must all be the same length, so $L$ must be chosen and fixed.
\item $N_{\rm circ}$ is the number of circuits in the bag for each region. This number can be chosen to be a constant (independent of $n$), as we argue below.
\item $N_{\rm con}$ is the number of random contexts in which each subcircuit will be tested.
\end{itemize}
First, we choose a bag of $N_{\rm circ}$ depth-$L$ subcircuits for each of the $M$ regions (see below for their construction).  Now, for each region $m\in[0\ldots M-1]$ and each of the subcircuits $\nu^m$ in that region's bag, we define $N_{\rm con}$ \emph{different} circuits that perform $\nu^m$ in different contexts, by choosing a subcircuit for each of the other regions at random from the corresponding bag, and performing all those subcircuits (including $\nu^m$) in parallel.  This circuit selection procedure is illustrated in \cref{fig:expts}.

This design ensures that (1) for each region, approximately $N_{\rm circ}$ different subcircuits are studied in detail, (2) each of these subcircuits is performed in $N_{\rm con}$ different contexts, and (3) those contexts vary independently across all the other regions.  

We have found that a small refinement improves the protocol in practice.  Often, the idle gate is less noisy than others, and the depth-$L$ idle circuit is the least noisy depth-$L$ circuit \emph{and} most sensitive to crosstalk.  We have found it useful to artificially boost the probability of sampling all-idle circuits when the random contexts are defined (\emph{not} when $\nu^m$ is drawn).  To do this, we sample context subcircuits normally, but replace each sample by the depth-$L$ idle circuit with probability $p_{\rm idle}$. This experiment design is also described by pseudocode in Appendix C. 

Two additional variations are useful in some circumstances.  First, in certain regimes, we find that crosstalk can be comprehensively detected without testing all $N_{\rm circ}$ subcircuits in each region's bag.  Iterating $\nu^m$ over a randomly chosen subset is sufficient.  Second, it is sometimes easier to \emph{sample} the subcircuits $\nu^m$ at random (with replacement) -- just as the contexts are sampled randomly -- than to iterate over them.  

The experiment defined above can be seen as a sparse filling of the hypercube defined by the exhaustive experiment, as long as $N_{\rm con}$ does not grow exponentially with $M$.  Ideally, it should not grow with $M$ at all.  In practice, we find a constant $N_{\rm con}$ (with respect to $M$) to be sufficient to detect crosstalk errors.  The protocol also requires specifying both $N_{\rm circ}$ and how to construct the subcircuits, which we discuss at the end of this section. The total number of experimental configurations to be performed for any fixed length $L$ is $N_{\rm exp} \approx M\times N_{\rm circ} \times N_{\rm con}$ \footnote{This is an approximately equal to and not a strict equality because any accidentally duplicate circuits generated during the sampling of contexts are removed.}, which scales linearly in the number of regions, $M = \mathcal{O}(n)$. 

Each of the circuits prescribed for this protocol should be repeated $N_{\rm rep}$ times to collect statistics. Each repetition yields a single datum, comprising a label for the circuit applied to each region ($\set_{r_i}$) and a bit string describing the measurement results from each region ($\res_{r_i}$).  This is a single sample from the distribution over settings and results $P(\Omega)$ that we seek to test for correlations that signal crosstalk errors.

As much as possible, the repetitions of the various circuits should be distributed uniformly over the entire time of the experiment -- \emph{not} performed all at once in a single chunk.  They may be \emph{rasterized} (each circuit is performed once, in succession, and this is repeated $N_{\rm rep}$ times), or \emph{randomized} (all the circuit repetitions are shuffled and performed in completely random order).  This minimizes the probability of systematic false positives caused by drift.  If behavior of the device (\eg error rates) is correlated with time -- \ie it drifts -- then if the settings are also correlated with time, this will produce spurious evidence of correlation between settings and results.  Time is an unobserved, or latent, variable; \eg in the simplest case an unobserved classical degree of freedom (\eg a two-state fluctuator) may cause drift by providing a fluctuating local potential. When a variable that is a common cause for multiple other variables is not observed, it can create a fictitious conditional dependence between these variables \cite{hernan_simpsons_2011}. Randomization and rasterization reduce or destroy correlations between the settings and time, reducing the risk of conflation; see also related discussion about conflation in \cref{sec:network_disc}. Finally, we note that rasterizing also facilitates concurrent drift detection with the same data \cite{proctor_detecting_2019}.

We saw above that the number of distinct experiments required to detect crosstalk errors for one partition of the QIP is $\mathcal{O}(n)$. This is multiplied by the number of partitions to get the overall cost of detecting crosstalk errors for the QIP. As discussed above, the number of required partitions for a QIP scales as $\mathcal{O}(n^2\log(n))$ in the worst case. Therefore the number of distinct experiments required for crosstalk detection with our protocol scales as $\mathcal{O}(n^3\log(n))$ (this reduces to $\mathcal{O}(n^2\log(n))$ if qubits are only coupled locally).

\subsubsection{Choosing the subcircuits for each region's bag}

Exactly what subcircuits to choose or define for each region is a critical component.  We have left it open for now for a simple reason:  there are many reasonable, yet very different, possible choices.  For the sake of concreteness, we specify particular circuits here.  But we also expect that new, creative, and perhaps objectively better choices can be usefully explored.

The subcircuits run on each region have two purposes:  to \emph{manifest} crosstalk, and to \emph{detect} crosstalk.  It may be that only certain circuits, when run on $r_i$, cause or amplify errors on $r_j$.  And it may also be that certain circuits on $r_j$ are more sensitive to these effects.  Our goal is to detect whatever crosstalk errors exist.  Therefore, in principle, the subcircuits chosen for each region's bag should be those that (1) cause the greatest effects on other regions, and (2) are the most sensitive to effects caused by other regions.

Given a specific physical model of crosstalk, it is possible to design subcircuits with these properties.  Or, given a \emph{parameterized} model of the sorts of crosstalk that might occur, it is possible to design a rather larger set of subcircuits that \emph{collectively} amplify all the effects appearing in that model.  (This is how the circuits for gate set tomography (GST) are chosen).

An entirely different approach is to switch from trying to detect \emph{all} low-weight crosstalk errors to focusing on the crosstalk errors that impact a specific application.  This motivates choosing subcircuits that are representative of the subroutines that appear in specific algorithms, and would emphasize detection of crosstalk errors that impact execution of those particular algorithms. There is no obvious general way to doing this since most algorithms will couple many qubits and not respect the region boundaries defined for crosstalk detection. However, if there are certain subcircuits, or \emph{circuit motifs}, that occur repeatedly in an algorithm or application, these can be used to define regions and subcircuits for crosstalk detection. It should be noted though that this approach will not detect crosstalk errors that might occur \emph{only} when these circuit motifs are all put together into an application circuit. Detecting such errors requires an application- or architecture-specific test.

But we have intentionally assumed no specific model and no specific application.  In the absence of any other guidance, \emph{random} circuits -- like those used in randomized benchmarking -- are a sensible choice.  These have certain known drawbacks; they are less sensitive than periodic circuits (\eg those used in GST or robust phase estimation) to some forms of noise because they twirl it \cite{dankert_exact_2009, gross_evenly_2007, blume-kohout_demonstration_2017}.  But random circuits are both common and hard to fool -- their sensitivity to noise is not always high, but it is reliable.  Therefore, we propose that the bag for each region be constructed by choosing $N_{\rm circ}$ subcircuits uniformly at random from an ensemble of random sequences of the processor's elementary gates.  The simulations presented in \cref{sec:sims} use this construction.

\subsection{Analyzing data to detect crosstalk errors}
\label{sec:impl}
The experiment described in \cref{sec:expt_design} generates data, which can be analyzed to detect and quantify crosstalk errors in a QIP. In this section we explain how to do this.

Running the circuits described above generates samples from a joint probability distribution of settings and results over the $M$ regions, $P(\set_{r_0}, \set_{r_1},...,\set_{r_{M-1}}, \res_{r_0}, \res_{r_1}, ..., \res_{r_{M-1}})$. 
Testing this joint distribution for violations of the conditions in \cref{eq:cond_ind} enables detecting whether where are crosstalk errors between any of the regions in the QIP. But we can go further, by determining the \emph{structure} of the crosstalk errors -- \ie which pairs of regions experience crosstalk errors. This can be achieved using techniques from causal inference that discover conditional dependence relationships between the $2M$ variables in this distribution. Specifically, we show how to adapt techniques developed to learn causal structure in Bayesian networks \cite{Neapolitan:2004vk, koller_probabilities_2009}, to detect the structure of crosstalk errors. 

A Bayesian network is a directed graph where each node represents a random variable and the edges represent joint probabilistic relationships between the variables. It is a concise representation of the joint distribution over the variables, with an edge indicating a conditional dependence between the variables -- an edge from node $i$ to node $j$ indicates that variable $j$ is dependent on variable $i$, when conditioned on the other variables in the graph. This is notated $(i \not\Perp j) ~|~ A$, where $A$ is a set representing all other nodes/variables in the graph.

Identifying causal network structure from data is an active and rapidly evolving area of research in the field of causal inference, and there are many algorithms available to do such causal network discovery \cite{spirtes_causal_2016}. These algorithms fall into two broad classes. 
The first, termed \emph{search-and-score methods}, enumerate or search through graph structures (each of which corresponds to a particular form of the joint distribution over the variables) and evaluates the how well each fits the data according to a score (which is often its likelihood \footnote{In practice one uses an information criterion as opposed to just the likelihood in order to avoid overfitting the data.}). 
The second class of algorithms, referred to as \emph{constraint-based methods}, operate by reconstructing a graph that is consistent with the conditional dependencies seen in the data by performing a series of hypothesis tests. Search-and-score methods are typically very computationally expensive, especially for datasets with a large number of variables,  so we focus on constraint-based algorithms in this work.

Any constraint-based algorithm for causal network structure discovery can be split into two parts: (i) a statistical test that tests for conditional independence (CI) of some sets of random variables, given samples and at a level of statistical significance, and (ii) a network discovery algorithm that repeatedly applies this CI test to determine the edges in the network. The keys to developing a ``good'' network learning algorithm are to formulate a CI test that is efficient and powerful, and to formulate a network discovery algorithm that is efficient, in the sense of needing to applying as few CI tests as possible. Given the mature body of research in this field, we seek to apply a previously developed network discovery algorithm to reveal the conditional dependence structure between the random variables we have in the context of crosstalk error detection. In the following subsections we present specific choices for the CI test and network discovery algorithm. 

We emphasize that although we are using tools traditionally used in causal inference, we are not making claims about causality. Specifically, an edge between nodes $\set_{r_i}$ and $\res_{r_j}$ (or $\res_{r_i}$ and $\res_{r_j}$) for $i\neq j$ does not imply a direct causal relationship between the regions $r_i$ and $r_j$, just that there is some crosstalk error between these regions. This is an important caveat. Even in the context of classical physics, it is well known that statistical causal discovery algorithms are only heuristics  for revealing causal relationships (especially in the presence of latent, or unobserved, variables) \cite{spirtes_causation_2000,koller_probabilities_2009,spirtes_causal_2016}. In quantum theory, even defining causality and a definite causal order between random variables is thorny \cite{brukner_quantum_2014, wood_lesson_2015}. 
So we emphasize that we are simply using causal inference tools to efficiently assess conditional independence relationships that form the basis of our model-free definition of crosstalk errors.

\subsubsection{Statistical tests for conditional independence}
\label{sec:condind_tests}
There are many statistical tests for conditional independence. In the protocol described in \cref{sec:expt_design} the random variables of interest represent experimental settings and measurement outcomes. Both are drawn from a finite set. Therefore, all random variables in a data set resulting from such experiments will be \emph{categorical}. For such variables, a well-motivated test for conditional independence is the log-likelihood ratio test, or $G^2$ test \cite{Bacciu:2013gm}. 

To describe the test statistic associated with this test, let us first describe the data. The dataset consists of samples from $K$ random variables $\mathbb{X} = \{\mathsf{X}_{k}\}_{k=0}^{K-1}$ some of which represent experimental settings ($\set_{r_i}$) and some of which represent measurement outcomes (the $\res_{r_i}$). We assume that each $\mathsf{X}_k$ takes values from a finite set $\mathbb{X}_k$ of size $|\mathbb{X}_k|$. Then the $G^2$ test statistic that tests for the conditional dependence between variable $\mathsf{X}_i$ and $\mathsf{X}_j$, conditioned on the variables in the set $A \subset \mathbb{X}$ is defined as \cite{Bacciu:2013gm}
\begin{widetext}
\begin{align}
	G^2(i,j ~|~ A) = 2 \sum_{x_i, x_j, x_A} n_{ijA}(x_i,x_j, x_A) ~ \log \frac{n_{ijA}(x_i,x_j,x_A)n_A(x_A)}{n_{iA}(x_i,x_A)n_{jA}(x_j,x_A)},
\end{align}
\end{widetext}
where $n_{ijA}(x_i, x_j, x_A)$ is the frequency of the random variables $(\mathsf{X}_i, \mathsf{X}_j, \mathsf{X}_A)$ taking on the values $(x_i, x_j, x_A)$ in the dataset, and similarly for the other quantities. Note that $\mathsf{X}_A$ is a composite random variable since one may want to condition on several variables, \ie $|A|>1$. Under the null hypothesis, where $(\mathsf{X}_i \Perp \mathsf{X}_j) ~|~A$, this test statistic is asymptotically distributed as chi-squared with degrees of freedom $df = (|\mathbb{X}_i|-1)(|\mathbb{X}_j| - 1)(|\mathbb{X}_A|)$. 
This test statistic is a scaled version of the empirical estimate of the conditional mutual information between variables $\mathsf{X}_i$ and $\mathsf{X}_j$, given $A$. Thus this quantity also has a convenient information theoretic interpretation \cite{Bacciu:2013gm}.

Finally, we note that in the simplest case where the conditioning set is null, $A=\emptyset $, this statistical test is often referred to as a \emph{homogeneity} or independence test (with $df=(|\mathbb{X}_i|-1)(|\mathbb{X}_j| - 1)$) since it tests whether the distribution of variable $i$ is the same (homogeneous) regardless of the value of the variable $j$. 

\subsubsection{Network discovery algorithms}
\label{sec:network_disc}
The second part of a constraint-based causal network structure learning algorithm applies a CI test on data to reconstruct a network consistent with the data. The PC algorithm by Spirtes and Glymour \cite{spirtes_algorithm_1991} is a popular network discovery algorithm that has been widely implemented and tested. Appendix \ref{app:pc} has a detailed description of the algorithm, but here we outline its basic steps. The PC algorithm starts with a complete undirected graph with edges between all nodes (each of which represents a variable in the dataset). Then each edge is tested for conditional independence, given some conditioning set $A$ comprising neighbors of the nodes connected by the edge, for conditioning sets of increasing size (starting from an empty set). The resulting undirected graph is called the \emph{skeleton}, and the last step applies certain edge orientation rules in order to estimate a directed acyclic graph (DAG) representing the causal relations in the data. 

For crosstalk error detection, we will omit the last, edge orientation, step of the PC algorithm and will focus on the graph skeleton. We do this because we are not interested in identifying causal relationships (for reasons mentioned at the beginning of the section) and simply wish to detect conditional dependence relationships that signal violation of the crosstalk-free model. 

In the worst case, the runtime of the PC algorithm grows exponentially with the number of variables. However, graph sparsity greatly reduces computational cost, and the algorithm has been demonstrated on data with hundreds and thousands of variables \cite{le_fast_2018}.
Furthermore, detecting crosstalk errors is simpler than general causal network learning, because we can enforce some sparsity by encoding physically motivated information into the graph from the start. For example, the edge between any two experimental settings can be removed if they are randomized according to the experiment design outlined in \cref{sec:expt_design}. 

The PC algorithm performs multiple hypothesis tests to determine conditional independence relationships between random variables. In such multiple hypothesis testing scenarios one typically applies a significance adjustment, such as the Bonferroni correction, to control the number of false positives (type-I errors). These corrections are not done in the standard PC algorithm, because controlling the family-wise error rate is complicated by the structure of the PC algorithm: one does not know how many hypothesis tests will be performed \emph{a priori}. However, we note that there have been recent attempts to incorporate statistical methods for controlling the false discovery rate by modifying the PC algorithm \cite{li_controlling_2009,strobl_estimating_2019}. Implementing this more complex algorithm may increase the reliability and statistical rigor of the crosstalk error detection protocol. Alternatively, $\alpha$-significance weak control of the family-wise error rate \footnote{Weak control of the family-wise error rate with a significance level of $\alpha$ means that the probability of rejecting one or more null hypothesis is at most $\alpha$ when all the null hypotheses are true. It is more common to demand strong control of the family-wise error rate at significance $\alpha$, meaning that the probability of rejecting one or more true null hypothesis is at most $\alpha$ regardless of which of the hypotheses are true.} can be maintained by setting the input significance of the standard algorithm to $\alpha/K$ where $K$ is the number of edges in the initial graph.

\subsubsection{Quantifying crosstalk errors}
\label{sec:quantifying}
Applying the PC algorithm to a dataset reporting the experimental settings and measurement outcomes for regions will reconstruct a graph whose edges can be used to detect crosstalk errors at a specified significance level. However, we can also use this analysis to statistically quantify the amount of crosstalk error across any edge that represents crosstalk. 

Let the edge that represents crosstalk in a reconstructed graph be between variables $\mathsf{X}$ and $\mathsf{Y}$, \ie $\mathsf{X} \rightarrow \mathsf{Y}$. Recall that $\mathsf{X}$ takes values in the set $\{x_0, ... x_{|\mathbb{X}|-1}\}$ and $\mathsf{Y}$ takes values in $\{y_0, ... y_{|\mathbb{Y}|-1}\}$. We compute the following total variation distance (TVD) estimates
\begin{align}
	&d^{\mathsf{X} \rightarrow \mathsf{Y}}_{ij} \nn \\ 
	&= \sum_{{z=y_0}}^{y_{|\mathbb{Y}|-1}} \Bigg| \frac{n_{xy}(x_i, z)}{\sum_{z=y_0}^{y_{|\mathbb{Y}|-1}}n_{xy}(x_i, z)} - \frac{n_{xy}(x_j, z)}{\sum_{z=y_0}^{y_{|\mathbb{Y}|-1}}n_{xy}(x_j, z)} \Bigg|, \nn
\end{align}
for $0 \leq i,j \leq |\mathbb{X}|-1$.
This quantity is a measure of the difference between the distribution of $\mathsf{Y}$ when $\mathsf{X}=x_i$ and when $\mathsf{X}=x_j$. We quantify the amount of crosstalk error across the edge $\mathsf{X}\rightarrow \mathsf{Y}$ as the maximum over all $i,j$, since this represents the maximum deviation in the distribution of $\mathsf{Y}$ when $\mathsf{X}$ is varied:
\begin{align}
\mathcal{C}_{\mathsf{X}\rightarrow \mathsf{Y}} = \max_{i,j} d_{ij}^{\mathsf{X}\rightarrow \mathsf{Y}}.
\end{align}
Often, we also calculate the median over these TVDs to understand how much of an outlier the maximum is.

One has to be a little careful with this definition when $Y$ is a result random variable and $X$ is a setting random variable. To see this, suppose $X=\set_1$ and $Y=\res_{2}$. Then, the most sensible thing is to compare the distributions of $\res_2$ generated by the same setting $\set_2$, as $\set_1$ is varied. This requires that $\set_2$ take on a value $s$ when $\set_1=i$ and $\set_2=j$ in the above definition. In this case, we calculate the above TVD for every such common setting $\set_2$ for a pair  $\set_1=i$ and $\set_2=j$, and maximize over these. If no such common settings exist (which can happen for example, in the experiment specified in \cref{sec:expt_design} since it is a randomized design), then we fail to compute a TVD for that edge.

\subsection{Discussion and limitations}
\label{sec:limitations}
The crosstalk error detection protocol developed in the previous subsections is efficient in terms of experiment number and has tractable post-processing complexity for QIPs with hundreds of qubits \cite{le_fast_2018}. However, we have made several assumptions and restrictions in order to obtain this efficiency. 
Our assumptions stem from the fact that we are targeting low-weight crosstalk errors, see \cref{sec:diverse}. This greatly restricts the set of realistic crosstalk errors, and we concentrate on detecting these. Here we discuss the implications of our assumptions, and the associated limitations of the protocol. 

Given a circuit layer in an $n$-qubit Markovian QIP that has crosstalk errors (or a family of layers for relative crosstalk), the question of whether our protocol will detect it or not is dictated by the following factors:
\begin{enumerate}
	\item The partition of the QIP into regions, since the protocol detects crosstalk errors across regions.
	\item Whether the particular layer(s) that exhibit the crosstalk error is (are) sampled in the lightweight experiment design.
	\item Whether the detection procedure, using a network discovery algorithm and pairwise conditional independence tests is sufficient to detect the error.
	\item Statistical power; do we collect enough samples to determine the signal from noise?
\end{enumerate}
In the following, we will discuss each of these in turn.

\textit{Factor 1:} 
We partition a QIP into regions based on the elementary operations in the device and this implies a poly($n$) number of necessary partitions. Such a partitioning ignores regions of larger size that are composed of $k>2$ qubits. Operations on such regions will be composed out of one- and two-qubit operations, and therefore any crosstalk error would still be generated by the elementary operations. The assumption behind ignoring partitions with such ``composite" regions is that any crosstalk error present between such regions will also be present between some set of regions composed of $k\leq 2$ qubits. 
%This is consistent with low-weight crosstalk errors under which if the dynamics of region $r_i$ are correlated with the settings on several regions, they are \emph{also} correlated with each of those regions' settings individually.

More fundamentally, a crosstalk error that exists between a region with $k-1$ qubits and another with one qubit, and does not exist when the $k-1$-qubit region is sub-partitioned in any way, must be a weight-$k$ crosstalk error. 
%Therefore, to restrict attention to poly($n$) possible partitions of an $n$-qubit QIP we must be satisfied with only reliably detecting crosstalk errors of weight $\leq k$ where $k$ is a constant, not scaling with $n$. Any crosstalk error with larger weight could be missed by the protocol. 
Given this, by choosing regions of size at most $2$ as suggested above, we are accepting that we have no guarantee of detecting crosstalk errors of weight 4 or higher.

%real-world crosstalk usually has pairwise manifestations, meaning that if the dynamics of region $r_i$ are correlated with the settings on several regions, they are \emph{also} correlated with each of those regions' settings individually.

\textit{Factor 2:} Since the lightweight experiment design is based on random sampling, $L$, $N_{\rm circ}$ and $N_{\rm con}$ dictate the probability that any particular layer will be present in the crosstalk detection experiment. There are an exponential number of possible layers -- if there are $g$ elementary gates that can be applied to each region in an $M$-region QIP, this results in $g^M$ possible layers that can be executed in this QIP. Therefore, in order to guarantee that any possible layer is included in the experiment with high probability, $L$ or $N_{\rm circ}$ are required to scale exponentially in $M$. Our lightweight experiment does not provide this guarantee. However, such a guarantee of including every possible layer should be unnecessary for realistic quantum computing architectures, where if crosstalk is present in one layer it is also present in many others since the source of the crosstalk is one or several of the operations within a layer, and these are present in exponentially many layers.
It is easy to imagine adversarial crosstalk error models that do not fit this bill -- \eg there are crosstalk errors on qubit 1 if and only if there is an $X_{\pi}$ gate is applied on \emph{all} the other qubits. Our protocol would almost certainly not detect this crosstalk error because of the low probability of sampling this particular layer. However, this is an adversarial error model that is high-weight (since the operation on qubit 1 is conditioned on the classical register recording the operation applied to all the other qubits). We sacrifice detecting such crosstalk errors in order to derive an efficient protocol. 
 %an extremely adversarial error model (which would require very non-local physics) that could \emph{only} be detected with an inefficient protocol. Hence, we restrict ourselves to detecting realistic crosstalk errors, which -- if present -- exhibit themselves in a large fraction of the possible layers, and hence are included in the circuits sampled in the lightweight experiment design.

\textit{Factor 3:} Using the PC algorithm to identify crosstalk structure in a QIP implies some subtle assumptions about the crosstalk errors. To clarify these, we first note that the PC algorithm is known to fail to detect causal network structure when the probability distribution being sampled from is not \emph{faithful} to the underlying causal graph \cite{spirtes_causation_2000, klimova_faithfulness_2015, spirtes_causal_2016}. In our context, faithfulness means that if there exists crosstalk between regions $r_i$ and $r_j$, then there exist at least some random variables in $r_i$ that exhibit dependence to some random variables $r_j$, vice versa, or both. The classic example \cite{koski_review_2012,klimova_faithfulness_2015} where the faithfulness assumption is violated and the PC algorithm fails is with three random variables $\sfx_1,\sfx_2,\sfx_3$, that are pairwise independent; \eg if $\sfx_1,\sfx_2, \sfx_3$ are binary, and $\sfx_3 = \sfx_1\oplus \sfx_2$. This means that $\sfx_i \Perp \sfx_j$, for any $i,j$, but $(\sfx_i \not\Perp \sfx_j) ~|~ \sfx_k$ (for $i\neq j\neq k$). 
So each pair $\sfx_i$ and $\sfx_j$ are \emph{conditionally dependent}, but \emph{marginally independent}.
 The PC algorithm's first step tests each pair of variables for marginal independence \cite{spirtes_causation_2000}. This step would indicate that all pairs are marginally independent, and therefore all edges would be removed and the algorithm would terminate. Therefore the PC algorithm evaluated on samples from this distribution (even in the infinite sample size limit) would yield a graph with three nodes and no edges, despite the fact that these variables are clearly dependent. An analogue of this example in the context of crosstalk detection in QIPs is the following: suppose one is trying to detect crosstalk caused by single qubit gates in an $n$-qubit QIP. The regions are composed of single qubits, and suppose that the crosstalk errors are such that with the circuits that are tested, one ends up preparing an entangled state of the $n$ qubits with any two-local marginal density matrix that is completely mixed (\eg multiparty data hiding states \cite{hayden_multiparty_2005}). Then the results of measuring any qubit will be uncorrelated with the results from any other qubit (if all other measurement results are ignored) and the PC algorithm would not indicate any crosstalk between regions. The basic problem is that the marginal/local states do not produce distributions over measurement outcomes that are faithful to the underlying dependence (and correlation) between local subsystems. Testing the dependence (or correlation) between a large number of subsystems would reveal strong dependence. But the PC algorithm orders its tests by increasing number of variables (increasing size of conditioning set) for efficiency, and declares two variables to be independent as soon as it fails to detect a dependence. Therefore, it would never perform the necessary tests to reveal the dependence, which is also the root cause of the failure in the pairwise independent, three variable example given above. 
 
 Fortunately, producing unfaithful distributions over the random variables in the crosstalk error detection setting appears to be extremely artificial. Every case where we have been able to manufacture such distributions requires either (i) high-weight crosstalk errors acting non-trivially on several regions, (ii) extremely large crosstalk errors (\eg errors causing $\frac{\pi}{2}$-rotations), or (iii) fine-tuned crosstalk that cancels or adds up in precise ways. Moreover, we have not encountered this issue in any of the physically-relevant crosstalk error models that we have simulated. Therefore, we note it as an issue to be aware of when using the PC algorithm, but something that does not seem to practically affect the performance of the crosstalk detection protocol developed here. Moreover, we note that the PC algorithm is not the only option for the graph discovery portion of the protocol; it is possible to apply other approaches to detect the conditional dependency relationships between the operational variables \cite{spirtes_causal_2016}, although we have not explored these.

\textit{Factor 4:} If the experiment design and data analysis technique are sufficient to detect the faulty layer, the statistical power of each of the hypothesis tests underlying the network discovery algorithm increases with $N_{\rm rep}$ (\ie the distribution of the test statistic narrow with sample size). However, as mentioned above, since the overall analysis involves an \emph{a priori} unknown number of hypothesis tests, it is difficult to estimate the detection accuracy of the whole procedure as a function of sample size. Hence, we simply advice as large an $N_{\rm rep}$ as possible to minimize statistical error.

\subsection{Guidance for selecting protocol parameters}
\label{sec:guidance}
This protocol has several user-adjustable parameters.  Their values can be chosen, but not arbitrarily -- they control the reliability and power of the experiment.  Here, we provide some heuristic guidance on how to choose them. 
\begin{enumerate}
	\item $N_{\rm rep}$ is the number of repetitions of each experiment. Increasing $N_{\rm rep}$ reduces statistical noise, at the cost of requiring more time to take data.  We suggest that this should be as large as possible, and no less than 1000.
	\item $L$ is the the length or depth of the circuits, 
and two useful rules of thumb suggest what $L$ should be.  The first is that longer circuits (all else being equal) can exhibit increased crosstalk effects and therefore permit more sensitive detection.  However, once $L$ becomes greater than $1/\epsilon$, where $\epsilon$ is the rate of stochastic errors or decoherence, generic noise tends to swamp the effect sought.  Therefore, $L$ should be as large as feasible, but no greater than $O(1/\epsilon)$.
	\item $N_{\rm circ}$ is the number of circuits in each region's bag, which in turn are randomly selected from the population of all depth-$L$ subcircuits on each region.  This parameter can be chosen to be a constant, independent of $M$, of order $10-30$.  At a minimum, it needs to be large enough to guarantee that all possible elementary gates that can be performed in a region appear in at least one of the subcircuits chosen for that region.
	\item $N_{\rm con}$ is the number of random contexts in which each subcircuit is intentionally performed.  Empirically, we find that it should be $\mathcal{O}(N_{\rm circs})$ -- but the best value for this parameter depends on the relative strength of crosstalk errors and local errors, which we refer to as \emph{signal-to-noise ratio}.  When crosstalk errors are comparable to local errors (\emph{low} signal-to-noise), we require $N_{\rm con} \sim N_{\rm circs}/2$. But if crosstalk errors dominate (\emph{high} signal-to-noise), we find that $N_{\rm con} \sim N_{\rm circs}/4$ is sufficient. 
	\item $p_{\rm idle}$ is the probability of sampling the length $L$ idle circuit on any of the $M-1$ regions when constructing a context. The recommended value of $p_{\rm idle}$ depends on whether the idle operation has a significantly lower local error rate than other operations. If it does, then we recommend choosing $p_{\rm idle}\sim 1/M$, so that there is probability $\sim (M-1)/M \approx 1$ for large $M$ that the idle circuit is among the contexts provided. Otherwise, $p_{\rm idle}$ should be smaller -- but even in this case, we find that  $p_{\rm idle}>0$ is often advantageous, although we do not have a good rule of thumb for how the optimal value varies with the idle error rate.
\end{enumerate}

As mentioned, the guidance for these parameters is based on empirical studies (except for the guidance for $L$, which is fairly standard in quantum benchmarking). It might be possible to develop more rigorous estimates for these parameters based on an analysis of the statistical convergence of the PC algorithm. However, the complexity of the PC algorithm makes analysis of its convergence difficult \cite{kalisch_estimating_2007, strobl_estimating_2019}, and hence we leave this as an avenue for potential future work.

is fairly standard, the guidance for $N_{\rm con}$ and $p_{\rm dile}$ is 

\section{Simulations}
\label{sec:sims}
In this section we illustrate the crosstalk error detection and quantification procedure developed above by simulation. The analysis of simulated data is performed using crosstalk error detection routines in the pyGSTi package \cite{noauthor_pygsti._nodate} that implement the PC algorithm as described above. 

\subsection{Two-qubit simulations}
\label{sec:two_qubit_sims}
We first simulate several kinds of crosstalk error on a two-qubit system, with qubits (which form the regions in this case) labeled 0 and 1. The settings for each qubit enumerate the subcircuits applied (which are just gate sequences in this case since each region is composed of a single qubit), and the experiments simulated correspond to the experiment design outlined in \cref{sec:expt_design}. In addition to the crosstalk error models, we also simulate local errors through a local depolarization channel (after every gate, including the idle) with depolarization rate $p_{\rm local}$. To illustrate the efficacy of the technique, in all the simulations below, we operate in the low signal-to-noise regime where the local error rates are comparable or larger than the crosstalk error rates. This is where we expect that it is most challenging to detect the crosstalk errors. 

The elementary one-qubit gates are assumed to be $X_{\pi/2}, Y_{\pi/2}, I$, where $I$ is an idle or identity gate that takes the same time as the other gates. The native state preparation is always ideally the $\ket{0}$ state for both qubits, and the measurements are in the computational basis. The gate sequences are determined according to the experiment design outlined in \cref{sec:expt_design}. In all of the simulated experiments, we follow the guidance in \cref{sec:guidance} and use the suggested value $N_{\rm con}=N_{\rm circs}/2$ (since the parameters chosen are in the low signal-to-noise regime). The values of $N_{\rm circs}, N_{\rm rep}, L$ and $p_{\rm idle}$ vary and are specified below for each case.

Finally, since the regions in this case are composed of single qubits, in this section we simplify notation and dispense with the additional $r$ subscript when denoting results and settings; \ie $\set_{r_i} \rightarrow \set_{i}$ and $\res_{r_i} \rightarrow \res_{i}$.

\begin{figure*}[t]
\centering
\stackunder[5pt]{\includegraphics[scale=0.5]{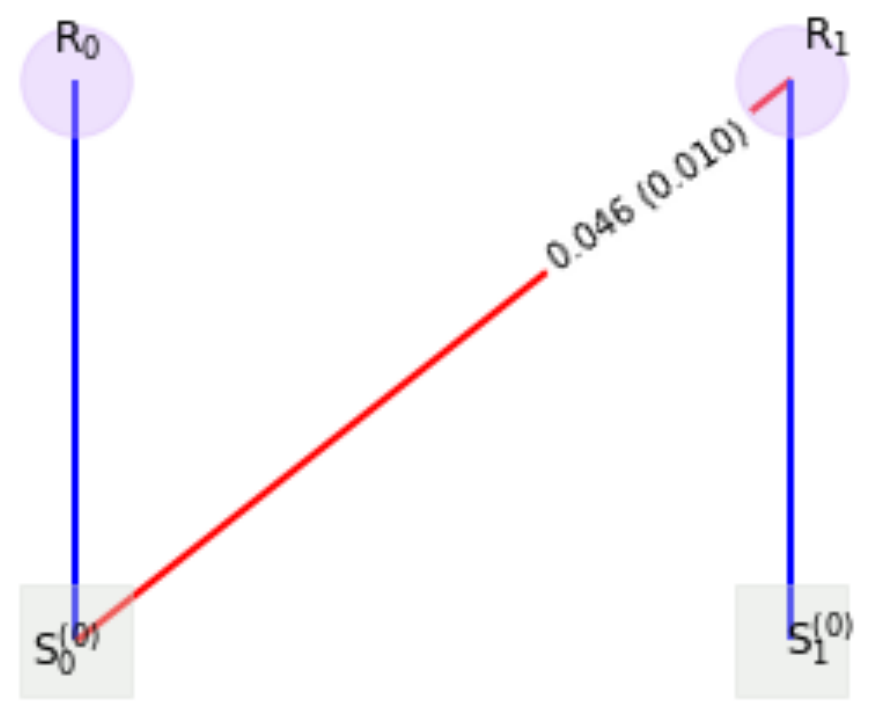}}{(a) Operation crosstalk error 1}
\hspace{0.5cm}
\stackunder[5pt]{\includegraphics[scale=0.5]{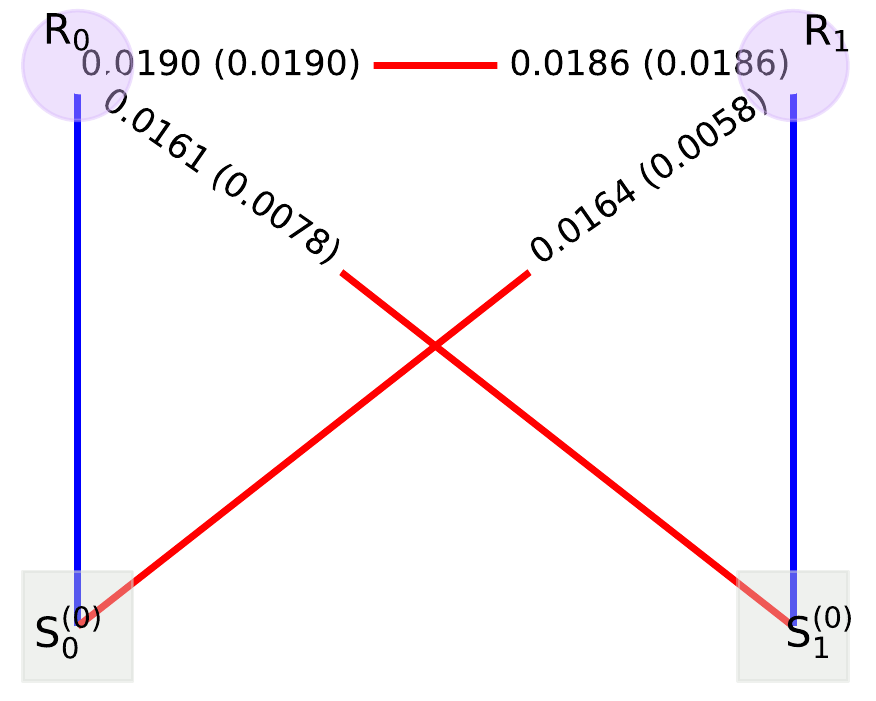}}{(b) Operation crosstalk error 2}
\hspace{0.5cm}
\stackunder[5pt]{\includegraphics[scale=0.5]{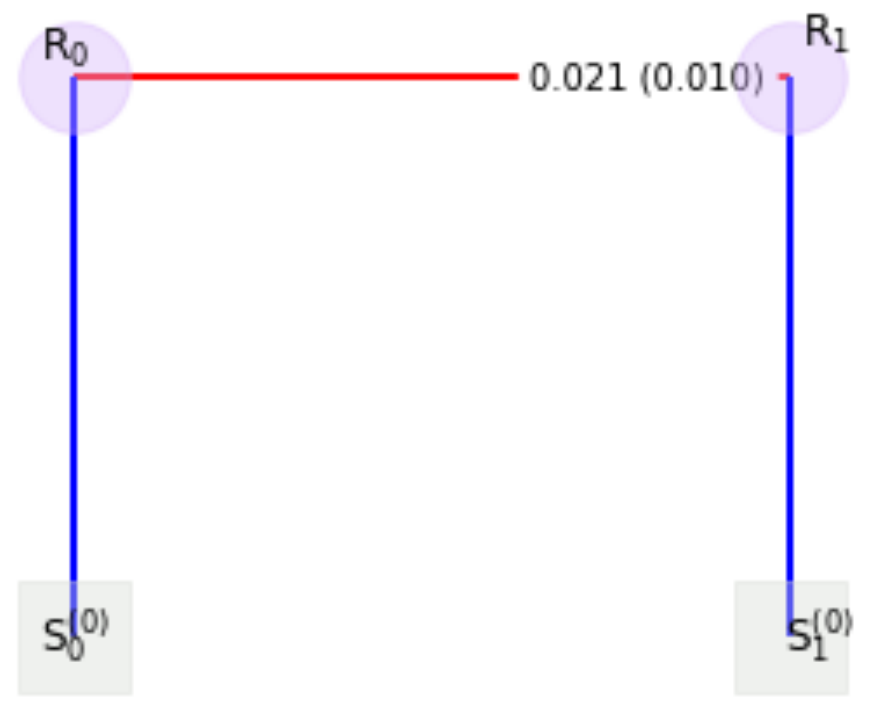}}{(c) Detection crosstalk error }
  \caption{Reconstructed graphs for various crosstalk error models in a systems of two qubits; see \cref{sec:two_qubit_sims} for details of error models. The regions in this case are composed on one qubit each. $\res_i$ represents the measurement result on qubit $i$ and $\set_i^{(0)}$ represents the setting on qubit $i$ (the superscript $(0)$ indexes the settings for a region -- in all our examples there is only one setting per region since only the applied gate sequence is varied). The blue edges indicate conditional dependencies between variables that are expected (\ie both variables belong to the same region). The red edges indicate conditional dependencies between variables in different regions, and these represent crosstalk. The red edges are labeled with the maximum TVD (and median TVD in parentheses) for that conditional dependence (see main text for definitions of these quantities). \label{fig:ct_graphs}}
\end{figure*}

\subsubsection{Operation crosstalk error 1}
\label{sec:classical_ct_sims}
The first error model we simulate is what we refer to as operation crosstalk error, and is also sometimes termed classical, or control line, crosstalk in literature. An $X_{\pi/2}$ gate on qubit 0 induces a depolarization channel on qubit 1 with depolarization rate $p$, \ie
\begin{align}
	\mathcal{X}_{\pi/2} \otimes \mathcal{I} \rightarrow \mathcal{X}_{\pi/2} \otimes \mathcal{D}_{p}, 
\end{align}
where $\mathcal{X}_{\pi/2}(\rho) = e^{-i \frac{\pi}{4}\sigma_x}\rho e^{i \frac{\pi}{4}\sigma_x}$ denotes a superoperator representation of a $X_{\pi/2}$ unitary rotation, $\mathcal{I}$ denotes an identity superoperator, and $\mathcal{D}_p(\rho) = (1-p)\rho + p I$ denotes a depolarization channel with depolarization probability $p$.

\cref{fig:ct_graphs}(a) shows the reconstructed crosstalk graph for this error model. The error model parameters used are: $p=10^{-2}, p_{\rm local}=10^{-2}$. The parameters defining the simulated experiment are $L=30$, $N_{\rm circs}=10, p_{\rm idle}=0.1, N_{\rm rep}=10^4$. The maximum number of unique circuits is  $N_{\rm exp}=M \times N_{\rm circs} \times \frac{N_{\rm circs}}{2} =100$. Finally, the significance level of the hypothesis tests used to test for conditional independence was set to $\alpha=0.01$. The red edge between settings in region 0 and results in region 1 in the graph signals the crosstalk between the qubits.

\subsubsection{Operation crosstalk error 2}
\label{sec:ham_ct_sims}
The next error model we simulate is an example of what is sometimes called coherent, or Hamiltonian, crosstalk. We model an $X_{\pi/2}$ gate on qubit 0 as inducing the desired rotation on qubit 0, but with an additional small two-qubit $Z\otimes Z$ Hamiltonian rotation as well, \ie
\begin{align}
	X_{\pi/2} \otimes I \rightarrow \exp\left(-\frac{i}{2}\left[\frac{\pi}{2} X\otimes I +\frac{\epsilon}{2} Z\otimes Z\right] \right).
\end{align}

\cref{fig:ct_graphs}(b) shows the reconstructed crosstalk graph for this error model. The error model parameters used are: $\epsilon=2\cdot10^{-2}, p_{\rm local}=10^{-2}$. The parameters defining the simulated experiment are $L=30, N_{\rm circs}=10, p_{\rm idle}=0, N_{\rm rep}=10^5$ (therefore, $N_{\rm exp}=100$), and the significance level of the hypothesis tests was set to $\alpha=0.01$. Note that the coherent crosstalk error shows up at $\sim \epsilon^2$ in the measurement probabilities since we are using random gate sequences, and this is why more samples are required to detect this crosstalk error.

The red edges in the crosstalk graph indicate crosstalk errors between the qubits. In this case there are conditional dependencies between settings and results in different regions, and \emph{also} between the results in different regions. There is no clear causal direction for this type of crosstalk error (and one can show using a model of this kind of crosstalk error and calculations such as in Appendix B that conditional dependencies between results are expected for this kind of crosstalk error).

\subsubsection{Detection crosstalk error}
The final error model we simulate is a model of crosstalk during the qubit measurement process. The measurement effects, indexed by the outcome values, are:
\begin{align}
	E_{00} &= \ket{00}\bra{00} \nn \\
	E_{01} &= \ket{01}\bra{01} \nn \\
	E_{10} &= (1-p_{\rm m})\ket{10}\bra{10} + p_{\rm m}\ket{11}\bra{11} \nn \\
	E_{11} &= (1-p_{\rm m})\ket{11}\bra{11} + p_{\rm m}\ket{10}\bra{10} \nn	
\end{align} 
In other words, if the measured value for the first qubit is $1$, there is a $p_{\rm m}$ probability that the measured value of the second qubit is flipped. This could, for example, model detection crosstalk due to scattered photons that flip the neighboring qubit state.
 
\cref{fig:ct_graphs}(c) shows the reconstructed crosstalk graph for this error model. The error model parameters used are: $p_{\rm m}=10^{-2}, p_{\rm local}=10^{-2}$. The parameters defining the simulated experiment are $L=10$, $N_{\rm circs}=20, p_{\rm idle}=0, N_{\rm rep}=10^5$ (therefore, $N_{\rm exp}=400$), and the significance level of the hypothesis tests was set to $\alpha=0.01$. Unlike the previous crosstalk error models, the effects of this error do not potentially build up over a gate sequence, and thus only impact the outcome probabilities weakly. Moreover, its effect is reduced in the experiments where the first qubit's outcome is 0 with high probability. For these reasons, we found that a larger $N_{\rm rep}$ and $N_{\rm circs}$ are required to detect this crosstalk error. 

The reconstructed graph in this case shows a red edge between the results on the two qubits indicating a conditional dependence that should not exist without some form of crosstalk error. 

\subsubsection{Crosstalk error quantification}
\begin{figure}[t]
\centering
\includegraphics[scale=0.5]{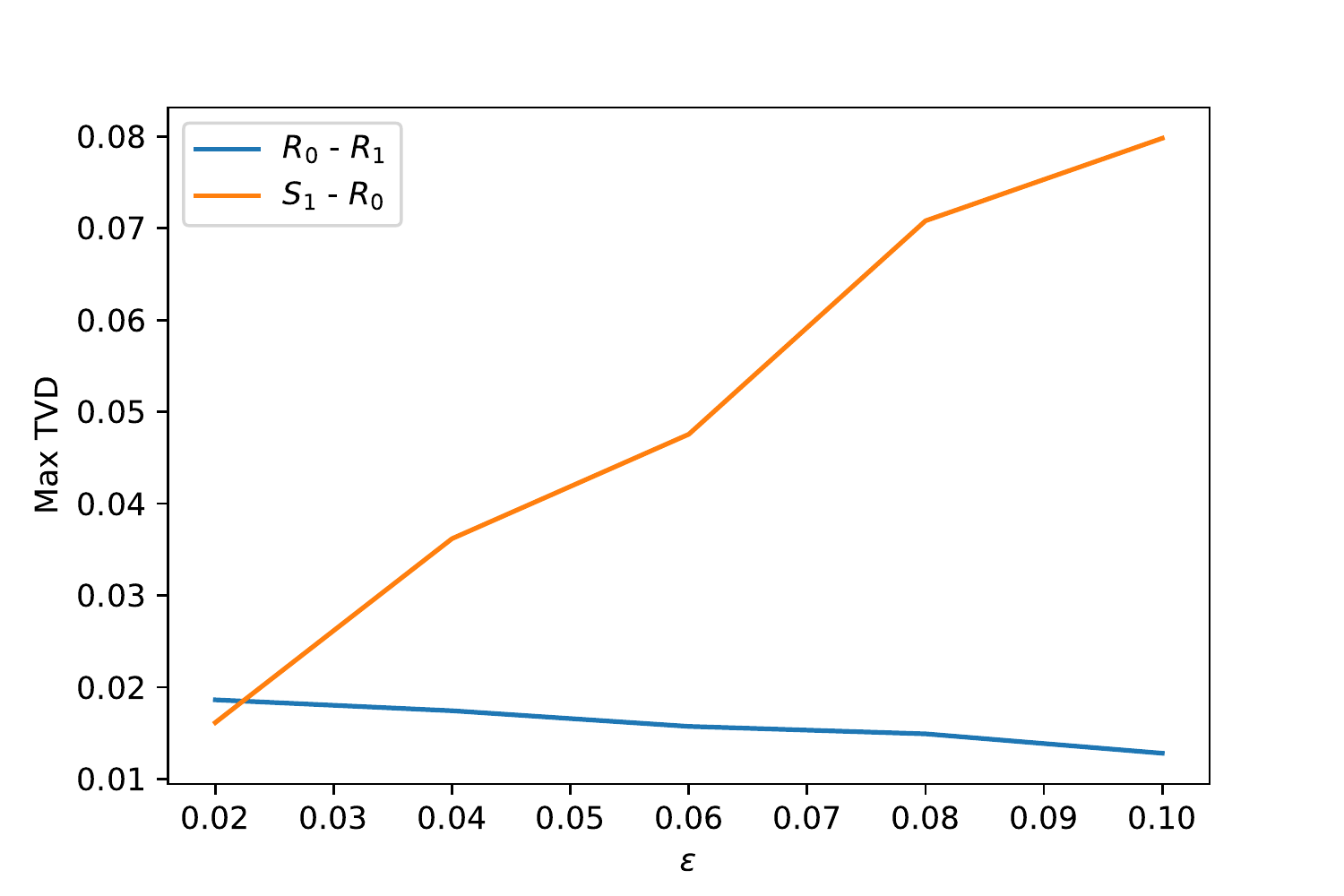}
  \caption{The weight of the edge from $\res_0$ to $\res_1$ and $\set_1$ to $\res_0$ versus the the physical crosstalk error magnitude, $\epsilon$, for the second operation crosstalk error model detailed in \cref{sec:ham_ct_sims}. \label{fig:edge_tvds}}
\end{figure}

It is important to keep in mind that the weights on the edges of a crosstalk graph are estimated maximum TVDs of outcome distributions, and not necessarily physical quantities like error rates. To illustrate this, in \cref{fig:edge_tvds} we return to the second operation crosstalk error model in \cref{sec:ham_ct_sims} and plot the weight of the edge from $\res_0$ to $\res_1$ and $\set_1$ to $\res_0$ as the amount of crosstalk, \ie the magnitude of the coherent $Z\otimes Z$ coupling term, is varied. The experiment sampling and physical parameters are the same as in \cref{sec:ham_ct_sims}, except that we use $N_{\rm circs}=5$. We see that while the max TVD increases (up to statistical variation) with increasing $\epsilon$ for one edge, it does not for the other. Even if the max TVD should vary monotonically with some crosstalk parameter when computed over all random circuits, in practice it is a function of the experiments that are sampled and therefore sensitive to finite sampling variations in these experiments. 

Therefore, the maximum TVD quantification should not be thought of as a direct measure of the \emph{physical degree} of crosstalk. It should instead be used as a way to identify the regions of a multiqubit device that require the most attention in terms of needing crosstalk mitigation. In addition, examining the experimental configuration that led to this maximum TVD often lends insight into the source of crosstalk errors.

\subsection{Six-qubit simulations}
\label{sec:six_qubit_sims}
\begin{figure*}[t]
\centering
\stackunder[5pt]{\includegraphics[scale=0.8]{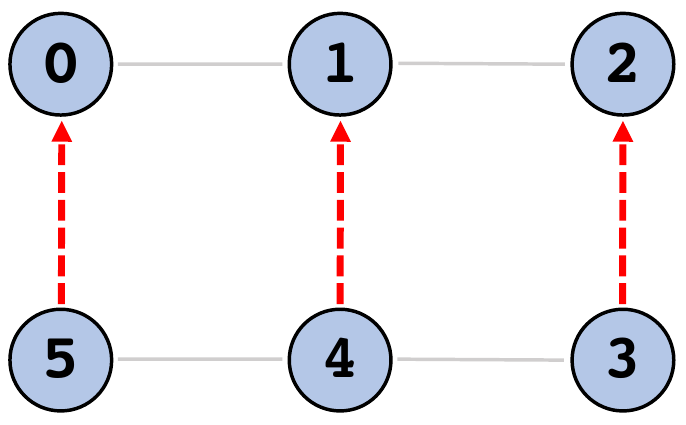}}{(a) 6 qubit QIP layout}
\hspace{2cm}
\stackunder[5pt]{\includegraphics[scale=0.5]{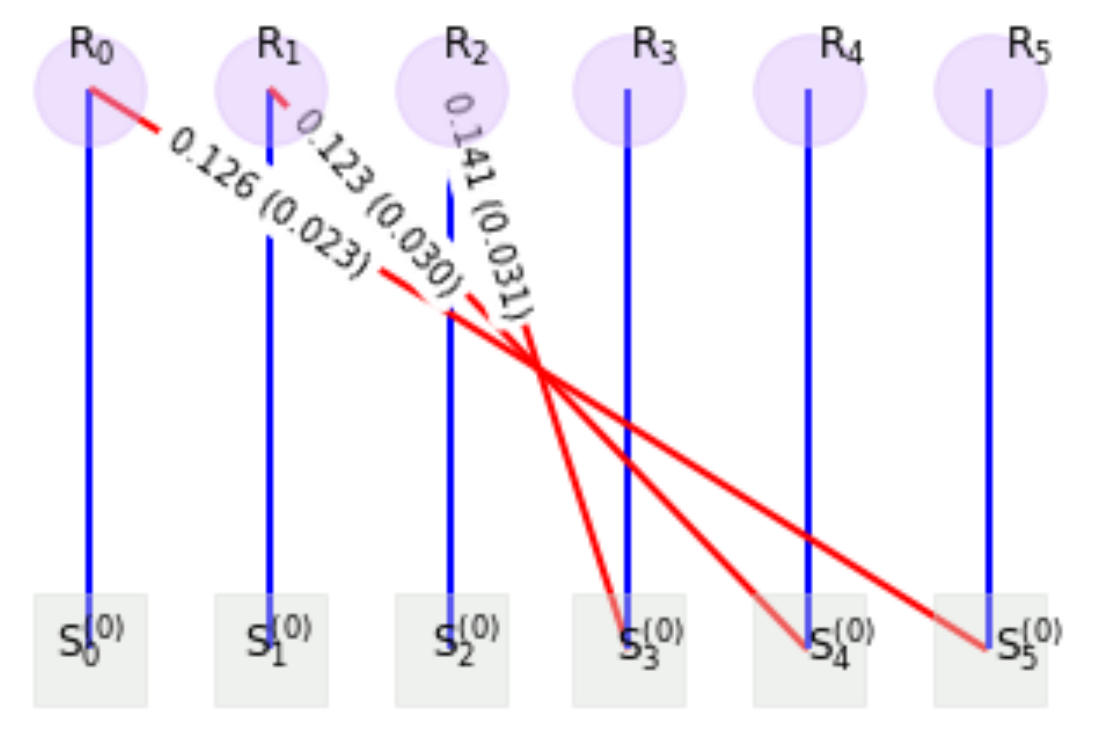}}{(b) Reconstructed crosstalk graph}
  \caption{Reconstructed graphs for 6 qubit QIP with operation crosstalk errors along vertical lines; see \cref{sec:six_qubit_sims} for details of error models. The regions in this case are composed on one qubit each. $\res_i$ represents the measurement results on qubit $i$ and $\set_i^{(0)}$ represents the setting on qubit $i$ (the superscript $(0)$ indexes the settings for a region -- in all our examples there is only one setting per region since only the gate sequence applied is varied). The blue edges indicate conditional dependencies between variables that are expected (\ie both variables belong to the same region). The red edges indicate conditional dependencies between variables in different regions, and this represent crosstalk. The red edges are labeled with the maximum TVD (and median TVD in parentheses) for that conditional dependence (see main text for definitions of these quantities).\label{fig:6q_ct_graphs}}
\end{figure*}
In this section we illustrate the scalability of the crosstalk error detection protocol by simulations on a 6-qubit device. The hypothetical device has a ladder layout, as shown in \cref{fig:6q_ct_graphs}(a) and we are interested in detecting the crosstalk errors caused by single qubit gates. So we partition the QIP into six regions with a single qubit in each. The settings for each qubit enumerate the gate sequences applied, and the experiments simulated correspond to the experiment design outlined in \cref{sec:expt_design}.

The crosstalk error model is similar to the first operation crosstalk error model detailed in \cref{sec:classical_ct_sims}; all single qubit gates on any of the qubits in the bottom line (qubits 3,4,5) result in a depolarizing channel with depolarization rate $p$ on the vertical neighboring qubit. In addition to these crosstalk errors we also simulate local errors through a depolarization channel with depolarization rate $p_{\rm local}$ after every gate and rate $p_{\rm idle}$ after every idle clock cycle. All other details (elementary gate set, state preparation and measurement, and form of experimental gate sequences used) are the same as in the two-qubit simulations.

 \cref{fig:6q_ct_graphs}(b) shows the reconstructed crosstalk graph for this error model. The parameters used were: $p=10^{-2}, p_{\rm local}=10^{-2}, p_{\rm idle}=5\times 10^{-3}, N_{\rm rep}=10^4$. The simulated experiment used $l=20$, $N_{\rm circs}=10$ (resulting in $N_{\rm exp}=300$), $p_{\rm idle}=0.1$, and the significance level of the hypothesis tests was set to $\alpha=0.01$. 
 
 The red edges in the crosstalk graph indicate crosstalk between the qubits that are vertical neighbors, as expected. We emphasize that this accurate crosstalk detection is achieved with just $300$ distinct experiments, which highlights the benefits of using a technique with experimental burden that scales essentially linearly with the number of qubits.

\section{Conclusions}
We make two contributions in this paper. First, we provide a universal and hardware-agnostic definition of crosstalk errors in terms of a model for QIP dynamics based on representations of gates, state preparations and measurements on the device. Second, we provide a model-free definition of crosstalk in terms of operational variables (QIP settings and measurement results), and develop a protocol for detecting crosstalk errors based on it.

The protocol is based on testing conditional independence relations between a potentially large number of random variables, and targets detection of low-weight crosstalk errors, which are a major concern for existing QIPs. We have tested the protocol and associated data processing on simulated experiments on QIPs with up to six qubits. The technique shows promise for crosstalk error detection on medium-scale QIPs since it requires a number of experiments that scales as $\mathcal{\tilde{O}}(n^3)$ in the worst-case, and scales as $\mathcal{\tilde{O}}(n^2)$ in realistic scenarios where qubit connectivity is limited. 

An avenue for future research is to explore the utility of alternatives to the PC algorithm for discovering the crosstalk structure in a QIP. The PC algorithm is arguably the most established constraint-based algorithm for causal network structure discovery, but there is an active field of study developing new approaches to causal network structure discovery, \eg the new kernel-based learning methods in Refs. \cite{Lopez-Paz:2015:TLT:3045118.3045273,mitrovic_causal_2018}, and it would be interesting to study whether any of these present any advantages when post-processing lightweight experimental data for crosstalk error detection. 

Of course, detecting crosstalk is just the first step. One would ideally like to also characterize crosstalk errors once detected in order to learn their form and possibly also their origin. In future work we will utilize the model-based definition of crosstalk developed here to construct efficient protocols for characterizing crosstalk errors.

\begin{acknowledgements}
We thank Sally Shrapnel for alerting us of the kernel-based causal network discovery methods developed in Refs. \cite{Lopez-Paz:2015:TLT:3045118.3045273,mitrovic_causal_2018}, and Michael Geller for pointing out the measurement crosstalk compensation performed in Refs. \cite{bialczak_quantum_2010, neeley_generation_2010, debnath_demonstration_2016, gong_genuine_2019, havlicek_supervised_2019}.
	Sandia National Laboratories is a multimission laboratory managed and operated by National Technology \& Engineering Solutions of Sandia, LLC, a wholly owned subsidiary of Honeywell International Inc., for the U.S. Department of Energy's National Nuclear Security Administration under contract {DE-NA0003525}. 
	This research was funded, in part, by the Office of the Director of National Intelligence (ODNI), Intelligence Advanced Research Projects Activity (IARPA) and by the  U.S.  Department  of  Energy, Office of Science, Office of Advanced Scientific Computing Research Quantum Testbed Program. This paper describes objective technical results and analysis. Any subjective views or opinions that might be expressed in the paper do not necessarily represent the views of the U.S. Department of Energy, IARPA, the ODNI, or the United States Government.  
\end{acknowledgements}

\bibliographystyle{plainurl}
\bibliography{crosstalk}

\begin{thebibliography}{10}

\bibitem{noauthor_pygsti._nodate}
{PyGSTi}. {A} python implementation of {Gate} {Set} {Tomography}.
\newblock URL: \url{http://www.pygsti.info/}.

\bibitem{ibmq-device-information_2019}
{IBM} {Q}iskit/ibmq-device-information/tenerife, 2019.
\newblock URL:
  \url{https://github.com/Qiskit/ibmq-device-information/tree/master/backends/tenerife/V1}.

\bibitem{Addis_2015}
Carole Addis, Francesco Ciccarello, Michele Cascio, G~Massimo Palma, and
  Sabrina Maniscalco.
\newblock Dynamical decoupling efficiency versus quantum non-markovianity.
\newblock {\em New Journal of Physics}, 17(12):123004, 2015.
\newblock \href {http://dx.doi.org/10.1088/1367-2630/17/12/123004}
  {\path{doi:10.1088/1367-2630/17/12/123004}}.

\bibitem{Bacciu:2013gm}
D~Bacciu, T~A Etchells, P~J~G Lisboa, and J~Whittaker.
\newblock {Efficient identification of independence networks using mutual
  information}.
\newblock {\em Computational Statistics}, 28:621, 2013.
\newblock \href {http://dx.doi.org/10.1007/s00180-012-0320-6}
  {\path{doi:10.1007/s00180-012-0320-6}}.

\bibitem{beals_quantum_2001}
Robert Beals, Harry Buhrman, Richard Cleve, Michele Mosca, and Ronald de~Wolf.
\newblock Quantum lower bounds by polynomials.
\newblock {\em Journal of the ACM}, 48(4):778--797, July 2001.
\newblock \href {http://dx.doi.org/10.1145/502090.502097}
  {\path{doi:10.1145/502090.502097}}.

\bibitem{bialczak_quantum_2010}
R.~C. Bialczak, M.~Ansmann, M.~Hofheinz, E.~Lucero, M.~Neeley, A.~D. O'Connell,
  D.~Sank, H.~Wang, J.~Wenner, M.~Steffen, A.~N. Cleland, and J.~M. Martinis.
\newblock Quantum process tomography of a universal entangling gate implemented
  with {Josephson} phase qubits.
\newblock {\em Nature Physics}, 6(6):409--413, 2010.
\newblock \href {http://dx.doi.org/10.1038/nphys1639}
  {\path{doi:10.1038/nphys1639}}.

\bibitem{blume-kohout_demonstration_2017}
Robin Blume-Kohout, John~King Gamble, Erik Nielsen, Kenneth Rudinger, Jonathan
  Mizrahi, Kevin Fortier, and Peter Maunz.
\newblock Demonstration of qubit operations below a rigorous fault tolerance
  threshold with gate set tomography.
\newblock {\em Nature Communications}, 8:1, 2017.
\newblock \href {http://dx.doi.org/10.1038/ncomms14485}
  {\path{doi:10.1038/ncomms14485}}.

\bibitem{PhysRevLett.103.210401}
Heinz-Peter Breuer, Elsi-Mari Laine, and Jyrki Piilo.
\newblock Measure for the degree of non-markovian behavior of quantum processes
  in open systems.
\newblock {\em Phys. Rev. Lett.}, 103:210401, 2009.
\newblock \href {http://dx.doi.org/10.1103/PhysRevLett.103.210401}
  {\path{doi:10.1103/PhysRevLett.103.210401}}.

\bibitem{brukner_quantum_2014}
Caslav Brukner.
\newblock Quantum causality.
\newblock {\em Nature Physics}, 10(4):259--263, 2014.
\newblock \href {http://dx.doi.org/10.1038/nphys2930}
  {\path{doi:10.1038/nphys2930}}.

\bibitem{buterakos_crosstalk_2018}
Donovan Buterakos, Robert~E. Throckmorton, and S.~Das~Sarma.
\newblock Crosstalk error correction through dynamical decoupling of
  single-qubit gates in capacitively coupled singlet-triplet semiconductor spin
  qubits.
\newblock {\em Physical Review B}, 97(4):045431, January 2018.
\newblock \href {http://dx.doi.org/10.1103/PhysRevB.97.045431}
  {\path{doi:10.1103/PhysRevB.97.045431}}.

\bibitem{chen_detector_2019}
Yanzhu Chen, Maziar Farahzad, Shinjae Yoo, and Tzu-Chieh Wei.
\newblock Detector tomography on {IBM} quantum computers and mitigation of an
  imperfect measurement.
\newblock {\em Physical Review A}, 100(5):052315, November 2019.
\newblock Publisher: American Physical Society.
\newblock \href {http://dx.doi.org/10.1103/PhysRevA.100.052315}
  {\path{doi:10.1103/PhysRevA.100.052315}}.

\bibitem{PhysRevB.70.195340}
W.~A. Coish and Daniel Loss.
\newblock Hyperfine interaction in a quantum dot: Non-markovian electron spin
  dynamics.
\newblock {\em Phys. Rev. B}, 70:195340, 2004.
\newblock \href {http://dx.doi.org/10.1103/PhysRevB.70.195340}
  {\path{doi:10.1103/PhysRevB.70.195340}}.

\bibitem{colombo_order-independent_2014}
Diego Colombo and Marloes~H Maathuis.
\newblock Order-{Independent} {Constraint}-{Based} {Causal} {Structure}
  {Learning}.
\newblock {\em J. Machine Learning Research}, 15:3921, 2014.
\newblock URL: \url{http://jmlr.org/papers/volume15/colombo14a/colombo14a.pdf}.

\bibitem{dankert_exact_2009}
Christoph Dankert, Richard Cleve, Joseph Emerson, and Etera Livine.
\newblock Exact and approximate unitary 2-designs and their application to
  fidelity estimation.
\newblock {\em Physical Review A}, 80(1):012304, 2009.
\newblock \href {http://dx.doi.org/10.1103/PhysRevA.80.012304}
  {\path{doi:10.1103/PhysRevA.80.012304}}.

\bibitem{davies_operational_1970}
E.~B. Davies and J.~T. Lewis.
\newblock An operational approach to quantum probability.
\newblock {\em Communications in Mathematical Physics}, 17(3):239--260, 1970.
\newblock \href {http://dx.doi.org/10.1007/BF01647093}
  {\path{doi:10.1007/BF01647093}}.

\bibitem{debnath_demonstration_2016}
S.~Debnath, N.~M. Linke, C.~Figgatt, K.~A. Landsman, K.~Wright, and C.~Monroe.
\newblock Demonstration of a small programmable quantum computer with atomic
  qubits.
\newblock {\em Nature}, 536(7614):63--66, August 2016.
\newblock \href {http://dx.doi.org/10.1038/nature18648}
  {\path{doi:10.1038/nature18648}}.

\bibitem{erhard_characterizing_2019}
Alexander Erhard, Joel~J. Wallman, Lukas Postler, Michael Meth, Roman Stricker,
  Esteban~A. Martinez, Philipp Schindler, Thomas Monz, Joseph Emerson, and
  Rainer Blatt.
\newblock Characterizing large-scale quantum computers via cycle benchmarking.
\newblock {\em Nature Communications}, 10(1):5347, 2019.
\newblock \href {http://dx.doi.org/10.1038/s41467-019-13068-7}
  {\path{doi:10.1038/s41467-019-13068-7}}.

\bibitem{Gambetta:2012ke}
Jay~M. Gambetta, A~D C{\'o}rcoles, S~T Merkel, B~R Johnson, John~A Smolin,
  Jerry~M Chow, Colm~A Ryan, Chad Rigetti, S~Poletto, Thomas~A Ohki, Mark~B
  Ketchen, and M~Steffen.
\newblock {Characterization of Addressability by Simultaneous Randomized
  Benchmarking}.
\newblock {\em Physical Review Letters}, 109(24):240504--5, 2012.
\newblock \href {http://dx.doi.org/10.1103/PhysRevLett.109.240504}
  {\path{doi:10.1103/PhysRevLett.109.240504}}.

\bibitem{gong_genuine_2019}
Ming Gong, Ming-Cheng Chen, Yarui Zheng, Shiyu Wang, Chen Zha, Hui Deng,
  Zhiguang Yan, Hao Rong, Yulin Wu, Shaowei Li, Fusheng Chen, Youwei Zhao,
  Futian Liang, Jin Lin, Yu~Xu, Cheng Guo, Lihua Sun, Anthony~D. Castellano,
  Haohua Wang, Chengzhi Peng, Chao-Yang Lu, Xiaobo Zhu, and Jian-Wei Pan.
\newblock Genuine 12-{Qubit} {Entanglement} on a {Superconducting} {Quantum}
  {Processor}.
\newblock {\em Physical Review Letters}, 122(11):110501, 2019.
\newblock \href {http://dx.doi.org/10.1103/PhysRevLett.122.110501}
  {\path{doi:10.1103/PhysRevLett.122.110501}}.

\bibitem{gross_evenly_2007}
D.~Gross, K.~Audenaert, and J.~Eisert.
\newblock Evenly distributed unitaries: {On} the structure of unitary designs.
\newblock {\em Journal of Mathematical Physics}, 48(5):052104, 2007.
\newblock \href {http://dx.doi.org/10.1063/1.2716992}
  {\path{doi:10.1063/1.2716992}}.

\bibitem{havlicek_supervised_2019}
Vojtech Havlicek, Antonio~D. Corcoles, Kristan Temme, Aram~W. Harrow, Abhinav
  Kandala, Jerry~M. Chow, and Jay~M. Gambetta.
\newblock Supervised learning with quantum-enhanced feature spaces.
\newblock {\em Nature}, 567(7747):209--212, 2019.
\newblock \href {http://dx.doi.org/10.1038/s41586-019-0980-2}
  {\path{doi:10.1038/s41586-019-0980-2}}.

\bibitem{hayden_multiparty_2005}
Patrick Hayden, Debbie Leung, and Graeme Smith.
\newblock Multiparty data hiding of quantum information.
\newblock {\em Physical Review A}, 71(6):062339, 2005.
\newblock \href {http://dx.doi.org/10.1103/PhysRevA.71.062339}
  {\path{doi:10.1103/PhysRevA.71.062339}}.

\bibitem{heinsoo_rapid_2018}
Johannes Heinsoo, Christian~Kraglund Andersen, Ants Remm, Sebastian Krinner,
  Theodore Walter, Yves Salathe, Simone Gasparinetti, Jean-Claude Besse, Anton
  Potocnik, Andreas Wallraff, and Christopher Eichler.
\newblock Rapid {High}-fidelity {Multiplexed} {Readout} of {Superconducting}
  {Qubits}.
\newblock {\em Physical Review Applied}, 10(3), 2018.
\newblock \href {http://dx.doi.org/10.1103/PhysRevApplied.10.034040}
  {\path{doi:10.1103/PhysRevApplied.10.034040}}.

\bibitem{hernan_simpsons_2011}
Miguel~A Hernan, David Clayton, and Niels Keiding.
\newblock The {Simpson}'s paradox unraveled.
\newblock {\em International Journal of Epidemiology}, 40(3):780--785, 2011.
\newblock \href {http://dx.doi.org/10.1093/ije/dyr041}
  {\path{doi:10.1093/ije/dyr041}}.

\bibitem{kalisch_estimating_2007}
Markus Kalisch and Peter Bühlmann.
\newblock Estimating {High}-{Dimensional} {Directed} {Acyclic} {Graphs} with
  the {PC}-{Algorithm}.
\newblock {\em Journal of Machine Learning Research}, 8(Mar):613--636, January
  2007.
\newblock URL: \url{http://www.jmlr.org/papers/v8/kalisch07a.html}.

\bibitem{klimova_faithfulness_2015}
Anna Klimova, Caroline Uhler, and Tamás Rudas.
\newblock Faithfulness and learning hypergraphs from discrete distributions.
\newblock {\em Computational Statistics \& Data Analysis}, 87:57--72, 2015.
\newblock \href {http://dx.doi.org/10.1016/j.csda.2015.01.017}
  {\path{doi:10.1016/j.csda.2015.01.017}}.

\bibitem{koller_probabilities_2009}
D~Koller and N~Friedman.
\newblock {\em Probabilistic {Graphical} {Models}}.
\newblock MIT Press, 2009.

\bibitem{koski_review_2012}
Timo~J.T. Koski and John Noble.
\newblock A {Review} of {Bayesian} {Networks} and {Structure} {Learning}.
\newblock {\em Mathematica Applicanda}, 40(1):51--103, 2012.
\newblock \href {http://dx.doi.org/10.14708/ma.v40i1.278}
  {\path{doi:10.14708/ma.v40i1.278}}.

\bibitem{Lad.Jel.etal-2010}
T~D Ladd, F~Jelezko, R~Laflamme, Y~Nakamura, C~Monroe, and J~L O'Brien.
\newblock {Quantum computers}.
\newblock {\em Nature}, 464:45, 2010.
\newblock \href {http://dx.doi.org/10.1038/nature08812}
  {\path{doi:10.1038/nature08812}}.

\bibitem{le_fast_2018}
Thuc~Duy Le, Tao Hoang, Jiuyong Li, Lin Liu, Huawen Liu, and Shu Hu.
\newblock A {Fast} {PC} {Algorithm} for {High} {Dimensional} {Causal}
  {Discovery} with {Multi}-{Core} {PCs}.
\newblock {\em IEEE/ACM Transactions on Computational Biology and
  Bioinformatics}, 16(5):1483--1495, 2019.
\newblock \href {http://dx.doi.org/10.1109/TCBB.2016.2591526}
  {\path{doi:10.1109/TCBB.2016.2591526}}.

\bibitem{li_controlling_2009}
Junning Li and Z~Jane Wang.
\newblock Controlling the {False} {Discovery} {Rate} of the
  {Association}/{Causality} {Structure} {Learned} with the {PC} {Algorithm}.
\newblock {\em J. Machine Learning Research}, 10:475, 2009.
\newblock URL: \url{http://www.jmlr.org/papers/v10/li09a.html}.

\bibitem{LI20181}
Li~Li, Michael~J.W. Hall, and Howard~M. Wiseman.
\newblock Concepts of quantum non-markovianity: A hierarchy.
\newblock {\em Physics Reports}, 759:1 -- 51, 2018.
\newblock \href {http://dx.doi.org/10.1016/j.physrep.2018.07.001}
  {\path{doi:10.1016/j.physrep.2018.07.001}}.

\bibitem{Lopez-Paz:2015:TLT:3045118.3045273}
David Lopez-Paz, Krikamol Muandet, Bernhard Sch\"{o}lkopf, and Ilya Tolstikhin.
\newblock Towards a learning theory of cause-effect inference.
\newblock In {\em Proceedings of the 32nd International Conference on
  International Conference on Machine Learning - Volume 37}, ICML'15, pages
  1452--1461, 2015.
\newblock URL: \url{http://dl.acm.org/citation.cfm?id=3045118.3045273}.

\bibitem{ma_dissipatively_2019}
Ruichao Ma, Brendan Saxberg, Clai Owens, Nelson Leung, Yao Lu, Jonathan Simon,
  and David~I. Schuster.
\newblock A dissipatively stabilized {Mott} insulator of photons.
\newblock {\em Nature}, 566(7742):51, February 2019.
\newblock \href {http://dx.doi.org/10.1038/s41586-019-0897-9}
  {\path{doi:10.1038/s41586-019-0897-9}}.

\bibitem{maciejewski_mitigation_2020}
Filip~B. Maciejewski, Zolt{\'{a}}n Zimbor{\'{a}}s, and Micha\l{} Oszmaniec.
\newblock Mitigation of readout noise in near-term quantum devices by classical
  post-processing based on detector tomography.
\newblock {\em Quantum}, 4:257, April 2020.
\newblock \href {http://dx.doi.org/10.22331/q-2020-04-24-257}
  {\path{doi:10.22331/q-2020-04-24-257}}.

\bibitem{majumdar_why_2018}
Rupak Majumdar and Filip Niksic.
\newblock Why is random testing effective for partition tolerance bugs?
\newblock {\em Proceedings of the ACM on Programming Languages}, 2(POPL):1--24,
  January 2018.
\newblock \href {http://dx.doi.org/10.1145/3158134}
  {\path{doi:10.1145/3158134}}.

\bibitem{mavadia_experimental_2018}
S.~Mavadia, C.~L. Edmunds, C.~Hempel, H.~Ball, F.~Roy, T.~M. Stace, and M.~J.
  Biercuk.
\newblock Experimental quantum verification in the presence of temporally
  correlated noise.
\newblock {\em npj Quantum Information}, 4(1):7, 2018.
\newblock \href {http://dx.doi.org/10.1038/s41534-017-0052-0}
  {\path{doi:10.1038/s41534-017-0052-0}}.

\bibitem{mazda_telecommunications_1993}
F~Mazda.
\newblock {\em Telecommunications {Engineer}'s {Reference} {Book}}.
\newblock Elsevier, 1993.
\newblock \href {http://dx.doi.org/10.1016/C2013-0-06529-2}
  {\path{doi:10.1016/C2013-0-06529-2}}.

\bibitem{mckay_three_2017}
David~C. McKay, Sarah Sheldon, John~A. Smolin, Jerry~M. Chow, and Jay~M.
  Gambetta.
\newblock Three-{Qubit} {Randomized} {Benchmarking}.
\newblock {\em Physical Review Letters}, 122(20):200502, 2019.
\newblock \href {http://dx.doi.org/10.1103/PhysRevLett.122.200502}
  {\path{doi:10.1103/PhysRevLett.122.200502}}.

\bibitem{mitrovic_causal_2018}
Jovana Mitrovic, Dino Sejdinovic, and Yee~Whye Teh.
\newblock Causal inference via kernel deviance measures.
\newblock In {\em Proceedings of the 32nd {International} {Conference} on
  {Neural} {Information} {Processing} {Systems}}, {NIPS}'18, pages 6986--6994,
  2018.
\newblock URL: \url{https://dl.acm.org/doi/10.5555/3327757.3327802}.

\bibitem{Neapolitan:2004vk}
Richard~E Neapolitan.
\newblock {\em {Learning Bayesian Networks}}.
\newblock Prentice Hall, 2004.

\bibitem{neeley_generation_2010}
Matthew Neeley, Radoslaw~C. Bialczak, M.~Lenander, E.~Lucero, Matteo
  Mariantoni, A.~D. O'Connell, D.~Sank, H.~Wang, M.~Weides, J.~Wenner, Y.~Yin,
  T.~Yamamoto, A.~N. Cleland, and John~M. Martinis.
\newblock Generation of three-qubit entangled states using superconducting
  phase qubits.
\newblock {\em Nature}, 467(7315):570--573, 2010.
\newblock \href {http://dx.doi.org/10.1038/nature09418}
  {\path{doi:10.1038/nature09418}}.

\bibitem{neill_blueprint_2017}
C.~Neill, P.~Roushan, K.~Kechedzhi, S.~Boixo, S.~V. Isakov, V.~Smelyanskiy,
  A.~Megrant, B.~Chiaro, A.~Dunsworth, K.~Arya, R.~Barends, B.~Burkett,
  Y.~Chen, Z.~Chen, A.~Fowler, B.~Foxen, M.~Giustina, R.~Graff, E.~Jeffrey,
  T.~Huang, J.~Kelly, P.~Klimov, E.~Lucero, J.~Mutus, M.~Neeley, C.~Quintana,
  D.~Sank, A.~Vainsencher, J.~Wenner, T.~C. White, H.~Neven, and J.~M.
  Martinis.
\newblock A blueprint for demonstrating quantum supremacy with superconducting
  qubits.
\newblock {\em Science}, 360(6385):195--199, April 2018.
\newblock \href {http://dx.doi.org/10.1126/science.aao4309}
  {\path{doi:10.1126/science.aao4309}}.

\bibitem{piltz_trapped-ion-based_2014}
C.~Piltz, T.~Sriarunothai, A.F. Varon, and C.~Wunderlich.
\newblock A trapped-ion-based quantum byte with $10^{-5}$ next-neighbour
  cross-talk.
\newblock {\em Nature Communications}, 5(1):4679, December 2014.
\newblock \href {http://dx.doi.org/10.1038/ncomms5679}
  {\path{doi:10.1038/ncomms5679}}.

\bibitem{preskill_quantum_2018}
John Preskill.
\newblock Quantum {Computing} in the {NISQ} era and beyond.
\newblock {\em Quantum}, 2:79, 2018.
\newblock \href {http://dx.doi.org/10.22331/q-2018-08-06-79}
  {\path{doi:10.22331/q-2018-08-06-79}}.

\bibitem{proctor_detecting_2019}
Timothy Proctor, Melissa Revelle, Erik Nielsen, Kenneth Rudinger, Daniel
  Lobser, Peter Maunz, Robin Blume-Kohout, and Kevin Young.
\newblock Detecting, tracking, and eliminating drift in quantum information
  processors.
\newblock July 2019.
\newblock arXiv: 1907.13608.
\newblock URL: \url{http://arxiv.org/abs/1907.13608}.

\bibitem{proctor_what_2017}
Timothy Proctor, Kenneth Rudinger, Kevin Young, Mohan Sarovar, and Robin
  Blume-Kohout.
\newblock What {Randomized} {Benchmarking} {Actually} {Measures}.
\newblock {\em Physical Review Letters}, 119(13), 2017.
\newblock \href {http://dx.doi.org/10.1103/PhysRevLett.119.130502}
  {\path{doi:10.1103/PhysRevLett.119.130502}}.

\bibitem{proctor_direct_2019}
Timothy~J. Proctor, Arnaud Carignan-Dugas, Kenneth Rudinger, Erik Nielsen,
  Robin Blume-Kohout, and Kevin Young.
\newblock Direct {Randomized} {Benchmarking} for {Multiqubit} {Devices}.
\newblock {\em Physical Review Letters}, 123(3):030503, 2019.
\newblock \href {http://dx.doi.org/10.1103/PhysRevLett.123.030503}
  {\path{doi:10.1103/PhysRevLett.123.030503}}.

\bibitem{reagor_demonstration_2018}
Matthew Reagor, Christopher~B. Osborn, Nikolas Tezak, Alexa Staley, Guenevere
  Prawiroatmodjo, Michael Scheer, Nasser Alidoust, Eyob~A. Sete, Nicolas
  Didier, Marcus P.~da Silva, Ezer Acala, Joel Angeles, Andrew Bestwick,
  Maxwell Block, Benjamin Bloom, Adam Bradley, Catvu Bui, Shane Caldwell,
  Lauren Capelluto, Rick Chilcott, Jeff Cordova, Genya Crossman, Michael
  Curtis, Saniya Deshpande, Tristan~El Bouayadi, Daniel Girshovich, Sabrina
  Hong, Alex Hudson, Peter Karalekas, Kat Kuang, Michael Lenihan, Riccardo
  Manenti, Thomas Manning, Jayss Marshall, Yuvraj Mohan, William O'Brien,
  Johannes Otterbach, Alexander Papageorge, Jean-Philip Paquette, Michael
  Pelstring, Anthony Polloreno, Vijay Rawat, Colm~A. Ryan, Russ Renzas, Nick
  Rubin, Damon Russel, Michael Rust, Diego Scarabelli, Michael Selvanayagam,
  Rodney Sinclair, Robert Smith, Mark Suska, Ting-Wai To, Mehrnoosh Vahidpour,
  Nagesh Vodrahalli, Tyler Whyland, Kamal Yadav, William Zeng, and Chad~T.
  Rigetti.
\newblock Demonstration of universal parametric entangling gates on a
  multi-qubit lattice.
\newblock {\em Science Advances}, 4(2):eaao3603, 2018.
\newblock \href {http://dx.doi.org/10.1126/sciadv.aao3603}
  {\path{doi:10.1126/sciadv.aao3603}}.

\bibitem{rudinger_probing_2019}
Kenneth Rudinger, Timothy Proctor, Dylan Langharst, Mohan Sarovar, Kevin Young,
  and Robin Blume-Kohout.
\newblock Probing {Context}-{Dependent} {Errors} in {Quantum} {Processors}.
\newblock {\em Physical Review X}, 9(2):021045, 2019.
\newblock \href {http://dx.doi.org/10.1103/PhysRevX.9.021045}
  {\path{doi:10.1103/PhysRevX.9.021045}}.

\bibitem{Sheldon:2016fh}
Sarah Sheldon, Easwar Magesan, Jerry~M Chow, and Jay~M. Gambetta.
\newblock {Procedure for systematically tuning up cross-talk in the
  cross-resonance gate}.
\newblock {\em Phys. Rev. A}, 93(6):060302, 2016.
\newblock \href {http://dx.doi.org/10.1103/PhysRevA.93.060302}
  {\path{doi:10.1103/PhysRevA.93.060302}}.

\bibitem{spirtes_introduction_2010}
Peter Spirtes.
\newblock Introduction to {Causal} {Inference}.
\newblock {\em Journal of Machine Learning Research}, 11:1643, 2010.
\newblock URL: \url{http://www.jmlr.org/papers/v11/spirtes10a.html}.

\bibitem{spirtes_algorithm_1991}
Peter Spirtes and Clark Glymour.
\newblock An {Algorithm} for {Fast} {Recovery} of {Sparse} {Causal} {Graphs}.
\newblock {\em Social Science Computer Review}, 9(1):67, 1991.
\newblock \href {http://dx.doi.org/10.1177/089443939100900106}
  {\path{doi:10.1177/089443939100900106}}.

\bibitem{spirtes_causation_2000}
Peter Spirtes, Clark~N. Glymour, and Richard Scheines.
\newblock {\em Causation, prediction, and search}.
\newblock MIT Press, Cambridge, Mass, 2nd ed edition, 2000.

\bibitem{spirtes_causal_2016}
Peter Spirtes and Kun Zhang.
\newblock Causal discovery and inference: concepts and recent methodological
  advances.
\newblock {\em Applied Informatics}, 3:3, 2016.
\newblock \href {http://dx.doi.org/10.1186/s40535-016-0018-x}
  {\path{doi:10.1186/s40535-016-0018-x}}.

\bibitem{strobl_estimating_2019}
Eric~V. Strobl, Peter~L. Spirtes, and Shyam Visweswaran.
\newblock Estimating and {Controlling} the {False} {Discovery} {Rate} of the
  {PC} {Algorithm} {Using} {Edge}-specific {P}-{Values}.
\newblock {\em ACM Transactions on Intelligent Systems and Technology},
  10(5):1--37, 2019.
\newblock \href {http://dx.doi.org/10.1145/3351342}
  {\path{doi:10.1145/3351342}}.

\bibitem{vandersypen_nmr_2005}
L.~M.~K. Vandersypen and I.~L. Chuang.
\newblock {NMR} techniques for quantum control and computation.
\newblock {\em Reviews of Modern Physics}, 76(4):1037--1069, January 2005.
\newblock \href {http://dx.doi.org/10.1103/RevModPhys.76.1037}
  {\path{doi:10.1103/RevModPhys.76.1037}}.

\bibitem{veitia_macroscopic_2017}
Andrzej Veitia, Marcus~P. da~Silva, Robin Blume-Kohout, and Steven~J. van Enk.
\newblock Macroscopic instructions vs microscopic operations.
\newblock 2017.
\newblock arXiv: 1708.08173.
\newblock URL: \url{http://arxiv.org/abs/1708.08173}.

\bibitem{Wallman_2014}
Joel~J Wallman and Steven~T Flammia.
\newblock Randomized benchmarking with confidence.
\newblock {\em New Journal of Physics}, 16(10):103032, 2014.
\newblock \href {http://dx.doi.org/10.1088/1367-2630/16/10/103032}
  {\path{doi:10.1088/1367-2630/16/10/103032}}.

\bibitem{1901.00267}
Adam Winick, Joel~J. Wallman, and Joseph Emerson.
\newblock Phenomenological measure of quantum non-markovianity.
\newblock 2019.
\newblock arXiv:1901.00267.
\newblock URL: \url{http://arxiv.org/abs/1901.00267}.

\bibitem{wood_lesson_2015}
Christopher~J Wood and Robert~W Spekkens.
\newblock The lesson of causal discovery algorithms for quantum correlations:
  causal explanations of {Bell}-inequality violations require fine-tuning.
\newblock {\em New Journal of Physics}, 17(3):033002, 2015.
\newblock \href {http://dx.doi.org/10.1088/1367-2630/17/3/033002}
  {\path{doi:10.1088/1367-2630/17/3/033002}}.

\end{thebibliography}

\clearpage
\widetext
\appendix

\section{Conditional versus marginal independence}
\label{sec:condind_marginalind}
\begin{figure}[h!]
\centering
  \includegraphics[scale=0.4]{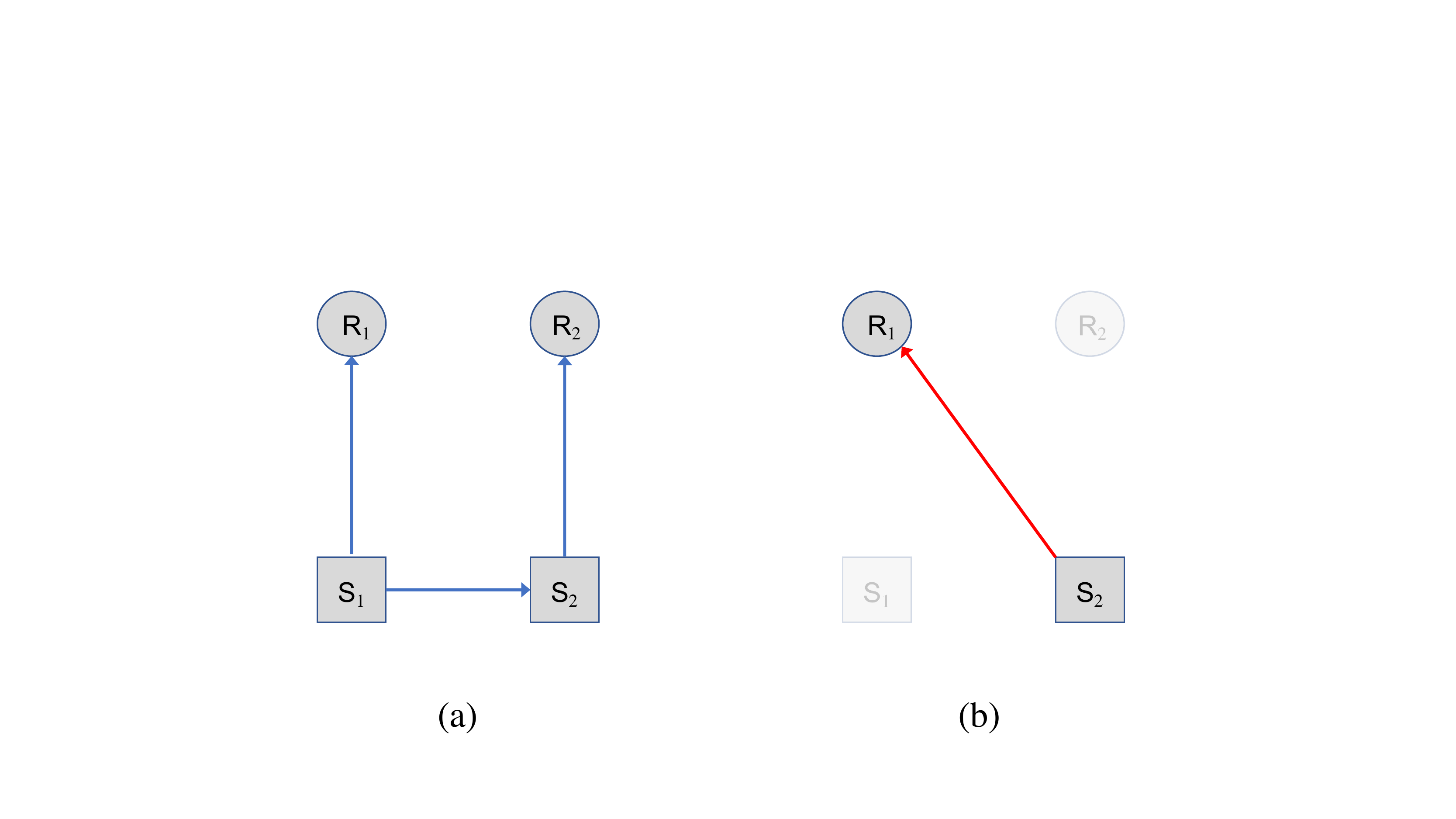}
  \caption{An example illustrating the difference between conditional and marginal independence between random variables. (a) shows the true causal relationships between the settings ($\set_1, \set_2$) and results ($\res_1, \res_2$) in a two qubit experiment. (b) shows the causal relationship inferred if a marginal independence test is performed on just variables $\res_1$ and $\set_2$. \label{fig:condind_vs_marginalind}}
\end{figure}

Our definition of crosstalk errors is based conditional independence of random variables. As mentioned in the main text, this is motivated by the central role played by conditional independence in defining causality in graphical models. In this Appendix we explain why we prefer to use the notion of conditional independence over marginal independence, which might seem simpler to work with. To do this, we refer to \cref{fig:condind_vs_marginalind}, which represents the random variables involved in a two qubit example: $\set_1, \set_2, \res_1, \res_2$. \cref{fig:condind_vs_marginalind}(a) represents the true dependence relationships between the variables. The arrow between $\set_1$ and $\set_2$ could be due to poor experiment design, whereby the settings on qubit 2 are not selected independently of those on qubit 1.

Excepting statistical issues (\ie given the underlying probability distribution over these random variables), the dependency relationships given by the graph in \cref{fig:condind_vs_marginalind}(a) would be reconstructed by examining conditional dependence relations. In contrast, if we only assess marginal independence between $\res_1$ and $\set_2$, the fact that both random variables have a common cause ($\set_1$) would create a fictitious dependence between these variables (see \cref{fig:condind_vs_marginalind}(b)). This is of course well known in statistics as the confounding of statistical association by an unobserved common cause \cite{hernan_simpsons_2011}.

Now, a marginal independence test would be sufficient if the experiment design was suitably randomized such that $\set_1$ and $\set_2$ are independent. However, for quantum computing platforms with many qubits a suitably randomized experiment design may be difficult to guarantee, and evaluation of association between random variables in distinct regions based on conditional independence testing is more reliable.

\section{Equivalence of two definitions of crosstalk}
\label{sec:condind_and_physics}
In the main text we presented two definitions of crosstalk-free QIPs, at two different layers of abstraction. The first, \textbf{Definition 1}, was stated in terms of properties of quantum operations (locality and independence) assuming a Markovian QIP, and the second, \textbf{Definition 2}, was stated operationally in terms of conditional independence between of classical random variables associated with experiments. In this Appendix we prove that these two definitions are in fact, equivalent; \ie A QIP is crosstalk-free according to \textbf{Definition 1} $\iff$ A QIP is crosstalk-free according to \textbf{Definition 2}.

\subsection{Conditional independence in terms of quantum operations}
Before we prove this equivalence we need to define the probabilities that arise in assessing conditional independence in terms of the quantum operations that form a model for the QIP.

We restrict ourselves to a QIP composed of two qubits for simplicity, and because it suffices to demonstrate the points we wish to make. In this case there are four random variables in the problem: $\res_i \in \{r_i^k\}_{k=1}^K, i=1,2$, the measurement results on the two qubits, and $\set_i \in \{r_i^l\}_{l=1}^L, i=1,2$, the settings on the two qubits. The settings will enumerate over some set of single-qubit gate sequences on each qubit.

Consider the conditional independence statement $P(\res_i | \set_i, \set_j, \res_j) =
 P(\res_i | \set_i)$ for $i,j\in \{1,2\}, i\neq j$, which captures the crosstalk-free condition between regions 1 and 2. What is this condition equivalent to in terms of physical states, operations and measurements? To address this question, we note that for any experiment, the Born rule dictates
 \begin{align}
 	P(r_1^i, r_2^j | s_1^k, s_2^l) = \Bra{E(r_1^i, r_2^j)} \mathcal{M}(s_1^k, s_2^l) \Ket{\rho_0},
 \end{align}
where $\Ket{E(r_1^i, r_2^j)}$ is a POVM indexed by the measurement results, $\mathcal{M}(s_1^k, s_2^l)$ is a general two qubit CPTP map indexed by the settings, and $\Ket{\rho_0}$ is the initial state of the two qubits. Note that we are using Hilbert-Schmidt representations of all of these quantities for notational simplicity. In fact, this is the only relation we can write down without making further assumptions about crosstalk, factorizability of operations, \emph{etc.} To obtain other probabilities we need to apply the usual conditioning and marginalization rules, \eg
\begin{align}
	P(r_1^i | s_1^k, s_2^l) &= \sum_n \Bra{E(r_1^i, r_2^n)} \mathcal{M}(s_1^k, s_2^l) \Ket{\rho_0} \nonumber \\
	P(r_1^i | s_1^k) &= \sum_n P(r_1^i | s_1^k, s_2^n) P(s_2^n | s_1^k) \nn \\
	&=\sum_n \sum_m \Bra{E(r_1^i, r_2^m)} \mathcal{M}(s_1^k, s_2^n) \Ket{\rho_0} ~P(s_2^n | s_1^k).
\end{align}

Using such relations we can write the crosstalk-free condition as
\begin{align}
P(\res_1 | \set_1, \set_2, \res_2)  &= P(\res_1 | \set_1 ) \nn \\
\Rightarrow  P(r_1^i | s_1^k, s_2^l, r_2^j) &= P(r_1^i | s_1^k)	\quad\quad \forall i,j,k,l \nn \\
\Rightarrow  \frac{P(r_1^i, r_2^j | s_1^k, s_2^l)}{P(r_2^j | s_1^k, s_2^l)} &= \sum_{n,m} P(r_1^i , r_2^n | s_1^k, s_2^m) P(s_2^m | s_1^k) \quad\quad \forall i,j,k,l \nn \\
\Rightarrow \Aboxed{ P(r_1^i, r_2^j | s_1^k, s_2^l) &= \left(\sum_z P(r_1^z, r_2^j | s_1^k, s_2^l)\right)\left(\sum_{n,m} P(r_1^i , r_2^n | s_1^k, s_2^m) P(s_2^m | s_1^k)\right) \quad\quad \forall i,j,k,l} 
\label{eq:condition} \\
\Rightarrow \Bra{E(r_1^i, r_2^j)} \mathcal{M}(s_1^k, s_2^l) \Ket{\rho_0} &= \left(\sum_z \Bra{E(r_1^z, r_2^j)} \mathcal{M}(s_1^k, s_2^l) \Ket{\rho_0}\right)\left(\sum_{n,m} \Bra{E(r_1^i, r_2^n)} \mathcal{M}(s_1^k, s_2^m) \Ket{\rho_0} P(s_2^m | s_1^k)\right) \nn \\
& \quad\quad\quad\quad\quad\quad \quad\quad \forall i,j,k,l \nn 
\end{align}
We can write a similar explicit equation for the condition $P(\res_2 | \set_1, \set_2, \res_1)  = P(\res_2 | \set_2 )$.

\subsection{Definition 1 $\Rightarrow$ Definition 2}
\label{sec:def1_to_def2}
A crosstalk-free Markovian QIP according to model-based definition, see \cref{sec:model}, has state preparations, gate operations, and measurements that satisfy:
\begin{align}
	\Ket{\rho_0} &= \Ket{\rho_0^1}\otimes \Ket{\rho_0^2} \nn \\
	\mathcal{M}(\set_1, \set_2) &= \mathcal{M}_1(\set_1)\otimes \mathcal{M}_2(\set_2) \nn \\
	\Ket{E(\res_1, \res_2)} &= \Ket{E(\res_1)}\otimes \Ket{E(\res_2})
	\label{eq:factorized}
\end{align}
We proceed to show that given such a model for operations, the QIP is also crosstalk-free under \textbf{Definition 2}; \ie $P(\res_i | \set_i, \set_j, \res_j) = P(\res_i | \set_i)$, for $i,j\in \{1,2\}, i\neq j$. To do so, we substitute the factorized forms in \cref{eq:factorized} into the explicit form of the condition $P(\res_1 | \set_1, \set_2, \res_2) = P(\res_1 | \set_1)$ given in \cref{eq:condition}:
\begin{align}
	P(r_1^i, r_2^j | s_1^k, s_2^l) &\mayeq \left(\sum_z P(r_1^z, r_2^j | s_1^k, s_2^l)\right)\left(\sum_{n,m} P(r_1^i , r_2^n | s_1^k, s_2^m) P(s_2^m | s_1^k)\right) \nn \\
	& \quad\quad\quad\quad\quad\quad \quad\quad \forall i,j,k,l \nn \\
	\Rightarrow \Bra{E(r_1^i)}\mathcal{M}_1(s_1^k)\Ket{\rho_0^1}\Bra{E(r_2^j)}\mathcal{M}_2(s_2^l)\Ket{\rho_0^2} &\mayeq \left(\sum_z \Bra{E(r_1^z)}\mathcal{M}_1(s_1^k)\Ket{\rho_0^1}\Bra{E(r_2^j)}\mathcal{M}_2(s_2^l)\Ket{\rho_0^2}\right) \cdot \nn \\
	& \quad\left( \sum_{n,m} \Bra{E(r_1^i)}\mathcal{M}_1(s_1^k)\Ket{\rho_0^1}\Bra{E(r_2^n)}\mathcal{M}_2(s_2^m)\Ket{\rho_0^2}P(s_2^m | s_1^k)\right) \nn \\
	& \quad\quad\quad\quad\quad\quad \quad\quad \forall i,j,k,l\nn \\
	\Rightarrow \Bra{E(r_1^i)}\mathcal{M}_1(s_1^k)\Ket{\rho_0^1}\Bra{E(r_2^j)}\mathcal{M}_2(s_2^l)\Ket{\rho_0^2} &\mayeq \Bra{E(r_2^j)}\mathcal{M}_2(s_2^l)\Ket{\rho_0^2}\Bra{E(r_1^i)}\mathcal{M}_1(s_1^k)\Ket{\rho_0^1}, 
\end{align}
which is obviously true. To arrive at the last line we have used the properties $\sum_z \Bra{E(r^z)}\mathcal{M}(s_k)\Ket{\rho_0}=1$ for any $k,\rho_0$, and $\sum_m P(s_2^m | s_1^k)=1$. We can verify that the equality $P(\res_2 | \set_2, \set_1, \res_1) = P(\res_2 | \set_2)$ also holds for quantum operations that satisfy \cref{eq:factorized}. This concludes the proof that the model-based definition of a crosstalk-free QIP implies our model-free definition of a crosstalk-free QIP.

\subsection{Definition 2 $\Rightarrow$ Definition 1}
To prove this direction, we will actually prove its contrapositive, namely, 
\begin{align}
	(\neg \textrm{locality}) \textrm{ or } (\neg \textrm{independence}) \Rightarrow  \textrm{violation of \textbf{Definition 2}}.
\end{align}
We proceed by showing that quantum operations that do not satisfy locality or independence lead to violations of the explicit form of $P(\res_1 | \set_1, \set_2, \res_2) = P(\res_1 | \set_1)$ stated in \cref{eq:condition}. The proofs straightforwardly generalize to the violation of $P(\res_2 | \set_2, \set_1, \res_1) = P(\res_2 | \set_2)$.

\subsubsection{Locality}
A Markovian QIP that does not satisfy the locality principle has state preparations, gate operations, or measurements (or all of these) that do not factorize as in \cref{eq:factorized}.

First consider the case where the initial state does not factorize, but all other operations do. Then, expanding \cref{eq:condition}, we get:
\begin{align}
	P(r_1^i, r_2^j | s_1^k, s_2^l) &\mayeq \left(\sum_z P(r_1^z, r_2^j | s_1^k, s_2^l)\right)\left(\sum_{n,m} P(r_1^i , r_2^n | s_1^k, s_2^m) P(s_2^m | s_1^k)\right) \nn \\
	& \quad\quad\quad\quad\quad\quad \quad\quad \forall i,j,k,l \nn \\
	\Rightarrow \Bra{E(r_1^i)} \otimes \Bra{E(r_2^j)} \mathcal{M}_1(s_1^k)\otimes \mathcal{M}_2(s_2^l)\Ket{\rho_0} &\mayeq \left(\sum_z\Bra{E(r_1^z)} \otimes \Bra{E(r_2^j)} \mathcal{M}_1(s_1^k)\otimes \mathcal{M}_2(s_2^l)\Ket{\rho_0}\right) \cdot \nn \\
	& \quad\left( \sum_{n,m} \Bra{E(r_1^i)} \otimes \Bra{E(r_2^n)} \mathcal{M}_1(s_1^k)\otimes \mathcal{M}_2(s_2^m)\Ket{\rho_0}P(s_2^m | s_1^k)\right) \nn \\
	& \quad\quad\quad\quad\quad\quad \quad\quad \forall i,j,k,l\nn \\
	\Rightarrow \Bra{E(r_1^i)} \otimes \Bra{E(r_2^j)} \mathcal{M}_1(s_1^k)\otimes \mathcal{M}_2(s_2^l)\Ket{\rho_0} &\mayeq \left( \Bra{E(r_2^j)} \Ket{\bar{\rho}_2(k,l)}\right) \Bra{E(r_1^i)} \Ket{\bar{\rho}_1(k)}\quad\quad \forall i,j,k,l,
\end{align}
where we have defined $\Ket{\bar{\rho}_1(k)}\equiv \tr_2\left(\mathcal{M}_1(s_1^k) \otimes \sum_m P(s_2^m | s_1^k)\mathcal{M}_2(s_2^m) \Ket{\rho_0}\right)$ and $\Ket{\bar{\rho}_2(k,l)}\equiv \tr_1\left(\mathcal{M}_1(s_1^k) \otimes \mathcal{M}_2(s_2^l) \Ket{\rho_0}\right)$. This last equality cannot be true in general since it is expressing a joint distribution over $r_1^i, r_2^j$ on the left hand side with a product of marginal distributions over these two variables on the right hand side. To see this more clearly, note that the equality must hold for all $i,j,k,l$, so suppose $\mathcal{M}_1(s_1^k)=\mathcal{M}_2(s_2^l)=\mathbf{1},~~\forall k,l$. In this case, the condition simplifies to:
\begin{align}
\Bra{E(r_1^i)} \otimes \Bra{E(r_2^j)} \Ket{\rho_0} &\mayeq \left( \Bra{E(r_2^j)} \Ket{\bar{\rho}_2}\right) \Bra{E(r_1^i)} \Ket{\bar{\rho}_1}\quad\quad \forall i,j,
\end{align}
where $\Ket{\bar{\rho}_{1(2)}}=\tr_{2(1)}\left( \Ket{\rho_0}\right)$. This equality cannot be true for any state $\rho_0$ that is not separable and therefore we conclude that conditional independence condition is violated in the case where the initial state does not factorize.

Next, consider the case where the gate operations do not factorize:
\begin{align}
	\exists a: \mathcal{M}(s_1^a, s_2^l) \neq \mathcal{M}_1(s_1^a)\mathcal{M}_2(s_2^l), ~~ \forall l,
\end{align}
\ie that the operation induced when the setting on the first qubit is $a$ is an entangling (non-factorizable) operation between the two qubits (regardless of what the setting is on qubit 1). We assume that the initial state and all measurement POVM elements factorize. 

Then, returning to the condition in \cref{eq:condition}, and considering the case $k=a$,
\begin{align}
	P(r_1^i, r_2^j | s_1^a, s_2^l) &\mayeq \left(\sum_z P(r_1^z, r_2^j | s_1^a, s_2^l)\right)\left(\sum_{n,m} P(r_1^i , r_2^n | s_1^k, s_2^m) P(s_2^m | s_1^k)\right) \nn \\
	& \quad\quad\quad\quad\quad\quad \quad\quad \forall i,j,l \nn \\
	\Rightarrow \Bra{E(r_1^i)} \otimes \Bra{E(r_2^j)} \mathcal{M}(s_1^a,s_2^l)\Ket{\rho^1_0}\otimes \Ket{\rho_0^2} &\mayeq \left(\sum_z\Bra{E(r_1^z)} \otimes \Bra{E(r_2^j)} \mathcal{M}(s_1^a,s_2^l)\Ket{\rho^1_0}\otimes \Ket{\rho_0^2}\right) \cdot \nn \\
	& \quad\left( \sum_{n,m} \Bra{E(r_1^i)} \otimes \Bra{E(r_2^n)} \mathcal{M}(s_1^a,s_2^m)\Ket{\rho^1_0}\otimes \Ket{\rho_0^2}P(s_2^m | s_1^k)\right)\nn \\
	& \quad\quad\quad\quad\quad\quad \quad\quad \forall i,j,l
	\label{eq:dummy_eq}
\end{align}
Now, suppose this equality holds. Then we can perform a weighted sum over $L$ on both sides to get
\begin{align}
\Bra{E(r_1^i)} \otimes \Bra{E(r_2^j)} \bar{\mathcal{M}}(s_1^a)\Ket{\rho^1_0}\otimes \Ket{\rho_0^2} =  \Bra{E(r_1^i)}\Ket{\bar{\rho}_1(a)}\Bra{E(r_2^j)}\Ket{\bar{\rho}_2(a)},  \quad\quad \forall i,j,\nn
\end{align}
where 
\begin{align}
\bar{\mathcal{M}}(s_1^a) &\equiv \sum_{l}	P(s_2^l|s_1^a)\mathcal{M}(s_1^a,s_2^l), \quad \quad \textrm{and} \nn \\
\Ket{\bar{\rho}_{1(2)}(a)} &\equiv \tr_{2(1)}\left(\bar{\mathcal{M}}(s_1^a)\Ket{\rho_0^1}\otimes \Ket{\rho_0^2}\right) \nn
\end{align}
Again, we have an expression with a joint distribution over $r_1^i, r_2^j$ on the left hand side and a product of marginals over the same variables on the right hand side. This cannot be true if $\mathcal{M}(s_1^a, s_2^l)$ does not factorize for all $L$, and therefore we conclude the original assumption of \cref{eq:dummy_eq} holding is false. Therefore, we have shown that the model-free crosstalk-free condition is violated when gate operations violate locality and do not factorize.

Finally, the proof that violation of locality in measurements -- \ie $\Ket{E(\res_1,\res_2} \neq \Ket{E(\res_1)}\otimes \Ket{E(\res_2)}$ -- results in a violation of the model-free crosstalk-free conditions follows straightforwardly from the corresponding proof for non-factorizable initial states since  state preparation and measurement are dual operations.

\subsubsection{Independence}
Now we proceed to show that even if locality is respected by a QIP, violations of independence in the model-based framework result in violations of the model-free definition of a crosstalk-free QIP. For simplicity we have assumed that the only settings correspond to gate operations, and we only have one choice for state preparation and measurement basis. Therefore violation of independence within a model that respects locality can only manifest in one way:
\begin{align}
\exists a,b \quad : \quad \mathcal{M}(s_1^a, s_2^b) &= \mathcal{E}_1(s_2^b)\mathcal{M}_1(s_1^a)\otimes \mathcal{M}_2(s_2^b).
\label{eq:indep_violation}
\end{align}
In other words, for some combination of settings, the operation done on the first qubit depends on the setting of the second qubit. Here, $\mathcal{E}_1()$ is some CPTP map on qubit 1.

Let us determine if this violation of independence results in a violation of the model-free condition in \cref{eq:condition}, by assuming all other operations respect locality and independence, and considering the case $k=a,l=b$:
\begin{align}
	P(r_1^i, r_2^j | s_1^a, s_2^b) &\mayeq \left(\sum_z P(r_1^z, r_2^j | s_1^a, s_2^b)\right)\left(\sum_{n,m} P(r_1^i , r_2^n | s_1^a, s_2^m) P(s_2^m | s_1^a)\right) \nn \\
	& \quad\quad\quad\quad\quad\quad \quad\quad  \forall i,j \nn \\
	\Rightarrow \Bra{E(r_1^i)} \mathcal{E}_1(s_2^b)\mathcal{M}_1(s_1^a) \Ket{\rho_0^1} \Bra{E(r_2^j)} \mathcal{M}_2(s_2^b)\Ket{\rho_0^2} &\mayeq \left(\sum_z\Bra{E(r_1^z)} \mathcal{E}_1(s_2^b)\mathcal{M}_1(s_1^a) \Ket{\rho_0^1} \Bra{E(r_2^j)} \mathcal{M}_2(s_2^b)\Ket{\rho_0^2}\right) \cdot \nn \\
	 \Bigg( \sum_{n} &  \Big[\Bra{E(r_1^i)} \mathcal{E}_1(s_2^b)\mathcal{M}_1(s_1^a) \Ket{\rho_0^1} \Bra{E(r_2^n)} \mathcal{M}_2(s_2^b)  \Ket{\rho_0^2}P(s_2^b | s_1^a)  \nn \\ 
	+ &
	 \sum_{m\neq b}  \Bra{E(r_1^i)} \mathcal{M}_1(s_1^a) \Ket{\rho_0^1} \Bra{E(r_2^n)} \mathcal{M}_2(s_2^m)\Ket{\rho_0^2}P(s_2^m | s_1^a)\Big] \Bigg) \nn \\
	& \quad\quad\quad\quad\quad\quad \quad\quad  \forall i,j \nn \\
	\Rightarrow \Bra{E(r_1^i)} \mathcal{E}_1(s_2^b)\mathcal{M}_1(s_1^a) \Ket{\rho_0^1} \Bra{E(r_2^j)} \mathcal{M}_2(s_2^b)\Ket{\rho_0^2} &\mayeq \Bra{E(r_2^j)} \mathcal{M}_2(s_2^b)\Ket{\rho_0^2} \cdot \nn \\
	 \Bigg[ \Bra{E(r_1^i)} \mathcal{E}_1(s_2^b)\mathcal{M}_1(s_1^a) \Ket{\rho_0^1} &P(s_2^b | s_1^a) + \Bra{E(r_1^i)} \mathcal{M}_1(s_1^a) \Ket{\rho_0^1} \left(1-P(s_2^b | s_1^a)\right)\Bigg] \quad\quad \forall i,j,\nn
\end{align}
where we have used the completeness property of the POVM elements to perform the sums over $z$ and $n$ in the last line. The equality on the last line holds if $\Bra{E(r_2^j)} \mathcal{M}_2(s_2^b)\Ket{\rho_0^2}=0$, so consider the cases where this quantity is non-zero (it cannot be zero for all $j$), and divide through both sides by this non-zero value. So for $j$ such that $\Bra{E(r_2^j)} \mathcal{M}_2(s_2^b)\Ket{\rho_0^2}\neq 0$, the condition we are evaluating becomes:
\begin{align}
	\Bra{E(r_1^i)} \mathcal{E}_1(s_2^b)\mathcal{M}_1(s_1^a) \Ket{\rho_0^1} \left(1-P(s_2^b | s_1^a)\right) &\mayeq  \Bra{E(r_1^i)} \mathcal{M}_1(s_1^a) \Ket{\rho_0^1} \left(1-P(s_2^b | s_1^a)\right) \quad\quad \forall i \nn \\
	\Rightarrow \Bra{E(r_1^i)} \mathcal{E}_1(s_2^b)\mathcal{M}_1(s_1^a) \Ket{\rho_0^1}  &\mayeq  \Bra{E(r_1^i)} \mathcal{M}_1(s_1^a) \Ket{\rho_0^1} \quad\quad \forall i,
\end{align}
where we have assumed $1-P(s_2^b|s_1^a)\neq 0$. Under what conditions is this last equality true when $\mathcal{E}_1(s_2^b)\neq\mathbf{1}$? The only other way for this equality to hold is if $\Bra{E(r_1^i)}\mathcal{E}_1(s_2^b) = \Bra{E(r_1^i)}, ~ \forall i$; \ie all the error maps act trivially on the measurement effects. Note that we could have written the violation of independence as a premultiplication error map (\ie $\mathcal{M}(s_1^a, s_2^b) = \mathcal{M}_1(s_1^a)\mathcal{E}_1(s_2^b)\otimes \mathcal{M}_2(s_2^b)$), in which case the equality would hold if the initial state is invariant under the error map. However, note that these CPTP maps represent the action of gate sequences, and the error map $\mathcal{E}_1(s_2^b)$ is the \emph{effective} error on qubit 1 when some sequence $\mathcal{M}_2(s_2^b)$ is performed on qubit 2, after the desired sequence on qubit 1, $\mathcal{M}_1(s_1^a)$ has been factored out. These sequences are composed of elementary gates, some subset of which violate the independence condition, which leads the whole map to violate the independence condition. However, for a sufficiently rich set of sequences if one sequence violates independence, then it is likely that others do as well (in other words, there are a set of $(a,b)$ satisfying \cref{eq:indep_violation}. And the probability that the measurement effects are invariant under all the effective error maps $\mathcal{E}_1(s_2^b)$ is extremely unlikely. Therefore we conclude that for a sufficiently rich set of settings, violation of independence results in violation of model-free definition of a crosstalk-free QIP.

\subsection{Definition 1 $\iff$ Definition 2}
The above subsections prove the two directions of implication required to establish equivalence between the model-based definition (\textbf{Definition 1}) and the model-free definition (\textbf{Definition 2}) of crosstalk-free QIPs.

\section{Pseudocode for lightweight experiment design}
\label{app:expt_design}

\begin{algorithm}[H]
\caption{Lightweight crosstalk detection experiment generation. The output is a set of roughly $M \times N_{\rm circs} \times N_{\rm con}$ experiments on an $M$ region QIP, with each experiment consisting of length $L$ circuits on each region.}
\begin{algorithmic}[1]
    \Procedure{CrosstalkExperiments}{$M, l, N_{\rm circs}, N_{\rm con}, p_{\rm idle}$}
      \For{ $0 < m < M-1$ } \Comment{Sample $N_{\rm circs}$ circuits for each region}
      	\State \texttt{bag}$_m$ $\leftarrow$ Sample of $N_{\rm circs}$ circuits length $L$, composed of elementary gates on region $m$
      \EndFor      
      \State \texttt{Expts} $\leftarrow$ \{\} \Comment{Initialize with empty list of experiments}
      \For{ $0 < m < M-1$ } 
      	\For{ $0 < n < N_{\rm circs}-1$ } \Comment{For each region, iterate over the $N_{\rm circs}$ circuits sampled for that region}
      		\State $s_m$ $\leftarrow$ Circuit number $n$ from \texttt{bag}$_m$
      		\For{ $0 < c < N_{\rm con}$ } \Comment{Generate $N_{\rm con}$ experiment with $s_m$ circuit on region $m$}
      			\For{ $0 < k < M-1$ }
      				\If{$k \neq m$}
      					\If{Unif(0,1) $< p_{\rm idle}$} \Comment{With probability $p_{\rm idle}$ region $k$ gets idle circuit}
      						\State $s_k$ $\leftarrow$ the length $L$ idle circuit on region $k$
      					\Else
      						\State $s_k$ $\leftarrow$ Sample a circuit from \texttt{bag}$_k$
      					\EndIf
      				\EndIf
      			\EndFor
      			\State Append to \texttt{Expts} the experiment defined by parallel application of $s_n$ (for $0<n<M-1$) to the $M$ regions
      		\EndFor
      	\EndFor
      \EndFor
      
      \State \texttt{Expts} $\leftarrow$ \textbf{RemoveDuplicates}(\texttt{Expts}) \Comment{Remove duplicate experiments}
      
      \State \textbf{return} \texttt{Expts}
    \EndProcedure
  \end{algorithmic}
\end{algorithm}

\section{Summary of the PC algorihtm}
\label{app:pc}
The PC algorithm is described in detail in references \cite{spirtes_causation_2000, colombo_order-independent_2014}, but we describe its main steps here, and comment on its implementation in the crosstalk detection context. 

Exhaustively checking for conditional dependence relations between $N$ data variables is exponential cost in terms of computational difficulty and required dataset size. To get around this, the PC algorithm uses insights from graph theory to perform a hierarchy of tests that can result in reduced costs, particularly in sparsely connected graphs (\ie datasets with sparse conditional dependence relations). The fundamental property exploited by the PC algorithm to reduce the number of conditional independency tests is this: two nodes $(X,Y)$ are conditionally independent given some subset of remaining nodes {\bf S}, if and only if they are conditionally independent given pa($X$) or pa($Y$), where pa($X$) are the parent nodes of $X$.

\Cref{alg:pc} presents pseudocode for the PC algorithm, adapted from Ref. \cite{spirtes_causation_2000}. Each variable is represented by a node in a graph $G$, and ${\bf Adj}(G,X)$ is the set of nodes adjacent to node $X$. The algorithm initializes by constructing the complete undirected graph with $N$ nodes. Then it prunes edges on this graph in a hierarchical manner: for every edge in the graph connecting nodes $(X,Y)$, it examines subsets of neighbors of one of the nodes of increasing size, $n$, (starting from the null set, $n=0$) and tests whether the two nodes are conditionally independent given the nodes in the subset. If so, it removes the edge. This procedure is repeated for every pair of connected nodes for increasing $n$, until no nodes have adjacency sets of size equal to or greater than $n$. At the end of this procedure, we have a pruned undirected graph with edges between nodes that are not conditionally independent under any conditioning set of adjacent nodes; this is often referred to as the \emph{skeleton} graph. Under the PC algorithm this undirected graph is passed to a subroutine OrientEdges that orients each edge in the graph using several orienting rules, resulting in a directed acyclic graph. We do not execute this portion of the algorithm for crosstalk detection and therefore do not provide details on that step. Interested readers are referred to \cite{spirtes_causation_2000,spirtes_algorithm_1991}.

\begin{algorithm}[H]
\caption{Pseudocode for the PC algorithm executing on a graph with $N$ nodes.  \label{alg:pc}}
\begin{algorithmic}[1]
    \Procedure{PC}{$N$}
    \State $G \leftarrow $ the complete undirected graph on a vertex set of size $N$
    \State $n \leftarrow 0$
    \Repeat
    	\Repeat
    		\State $(X,Y) \leftarrow $ an ordered pair of variables that are adjacent in $G$ such that {\bf Adj}($G, X) \backslash \{Y\}$ has cardinality $\geq n$
    		\State {\bf S} $\leftarrow$ a subset of {\bf Adj}($G, X) \backslash \{Y\}$ of cardinality $n$
    		\If{$(X,Y)$ are conditional independent given {\bf S}}
    			\State delete edge $X-Y$ from $G$
    			\State add {\bf S} to {\bf SepSet}($X$) and {\bf SepSet}($Y$)
    		\EndIf
    	\Until all ordered pairs of adjacent variables $(X,Y)$ such that $|${\bf Adj}($G, X) \backslash \{Y\}|\geq n$ and all {\bf S}$\subset $ {\bf Adj}($G, X) \backslash \{Y\}$ such that $|{\bf S}|=n$ have been tested for conditional independence. 
    	\State $n \leftarrow n+1$
    \Until for each order pair of adjacent vertices $(X,Y)$, $|{\bf Adj}(G, X) \backslash \{Y\}| < n$
    \State $G_{\rm orient} \leftarrow $ OrientEdges($G$, {\bf SepSet}($G$))
    \State \textbf{return} $G_{\rm orient}$
    \EndProcedure
  \end{algorithmic}
\end{algorithm}

The PC algorithm is stated in \cref{alg:pc} in terms of abstract conditional independence tests. These are implemented statistically in most implementations. Also, one might be concerned that the order in which pairs of variables are considered will effect the resulting undirected graph, and indeed this is true. However, one can formulate an order-independent version of the PC algorithm \cite{colombo_order-independent_2014} that removes this issue, and this is the version we use for crosstalk detection.

\end{document}